\pdfoutput=1
\documentclass[fleqn,usenatbib]{mnras}

\usepackage{newtxtext,newtxmath}

\usepackage[T1]{fontenc}

\usepackage{ae,aecompl}

\usepackage{graphicx}	
\usepackage{amsmath}	
\usepackage{amssymb}	
\usepackage{enumitem}
\usepackage{csvsimple}
\usepackage{multirow}
\usepackage{siunitx} 
\usepackage[utf8]{inputenc}
\usepackage{dirtytalk}
\usepackage{soul}
\usepackage{xcolor}

\hypersetup{pdfauthor={J. den Brok},
               pdftitle={Probing the AGN Unification Model at redshift z $\sim$ 3 with MUSE observations of giant Ly$
               		    \alpha$ nebulae},
               pdfkeywords={Galaxy: active, Galaxy: high-redshift, Quasar: general,  Quasar: emission lines, Intergalactic medium},
               bookmarksnumbered=true}

\newcommand{  \hei      }{\ifmmode {\rm He}\,\textsc{i} \else He\,\textsc{i}\fi}
\newcommand{  \heii     }{\ifmmode {\rm He}\,\textsc{ii} \else He\,\textsc{ii}\fi}
\newcommand{  \HeIIuv   }{\ifmmode {\rm He}\,\textsc{ii}\,\lambda1640 \else He\,\textsc{ii}\,$\lambda1640$\fi}
\newcommand{  \HeIIop   }{\ifmmode {\rm He}\,\textsc{ii}\,\lambda4686 \else He\,\textsc{ii}\,$\lambda4686$\fi}

\title[MUSE: Extended Ly$\alpha$ Emission morphology]{
Probing the AGN Unification Model at redshift z $\sim$ 3 with MUSE observations of giant Ly$\alpha$ nebulae
}

\author[J.~S. den Brok et al.]{
Jakob S. den Brok,$^{1,2}$\thanks{E-mail: jdenbrok@astro.uni-bonn.de}
S. Cantalupo,$^{1}$
R. Mackenzie, $^{1}$
R.~A. Marino,$^{1}$
G. Pezzulli$^{1}$  
\newauthor
J.~Matthee,$^{1}$ 
S.~D. Johnson,$^{3,4}$
M. Krumpe,$^{5}$
T. Urrutia,$^{5}$
W. Kollatschny$^{6}$
\\
$^{1}$Department of Physics, ETH Z\"urich, Wolfgang-Pauli-Strasse 27, 8093, Z\"urich,  Switzerland\\
$^{2}$Argelander-Institut für Astronomie, Universität Bonn, Auf dem Hügel 71, 53121 Bonn, Germany\\
$^{3}$Department of Astrophysical Sciences, 4 Ivy Lane, Princeton University, Princeton, NJ 08544, USA \\
$^{4}$The Observatories of the Carnegie Institution for Science, 813 Santa Barbara Street, Pasadena, CA 91101, USA\\ 
$^{5}$Leibniz-Institut f\"ur Astrophysik Potsdam (AIP), An der Sternwarte 16, D-14482 Potsdam, Germany\\
$^{6}$ Institut f\"ur Astrophysik, Universit\"at G\"ottingen, Friedrich-Hund Platz 1, 37077, G\"ottingen, Germany
}

\date{Accepted XXX. Received YYY; in original form ZZZ}

\pubyear{2019}

\begin{document}
\label{firstpage}
\pagerange{\pageref{firstpage}--\pageref{lastpage}}
\maketitle

\begin{abstract}




\noindent A prediction of the classic active galactic nuclei (AGN) unification model is the presence of ionisation cones with different orientations depending on the AGN type. Confirmations of this model exist for present times, but it is less clear in the early Universe. Here, we use the morphology of giant Ly$\alpha$ nebulae around AGNs at redshift z$\sim$3 to probe AGN emission and therefore the validity of the AGN unification model at this redshift. We compare the spatial morphology of 19 nebulae previously found around type I AGNs with a new sample of 4 Ly$\alpha$ nebulae detected around type II AGNs. Using two independent techniques, we find that nebulae around type II AGNs are more asymmetric than around type I, at least at radial distances $r>30$~physical kpc (pkpc) from the ionizing source. We conclude that the type I and type II AGNs in our sample show evidence of different surrounding ionising geometries. This suggests that the classical AGN unification model is also valid for high-redshift sources. Finally, we discuss how the lack of asymmetry in the inner parts (r$\lesssim$30 pkpc) and the associated high values of the \heii\ to Ly$\alpha$ ratios in these regions could indicate additional sources of (hard) ionizing radiation originating within or in proximity of the AGN host galaxies. This work demonstrates that the morphologies of giant Ly$\alpha$ nebulae can be used to understand and study the geometry of high redshift AGNs on circum-nuclear scales and it lays the foundation for future studies using much larger statistical samples. 



\end{abstract}




\noindent

 
    
    
    

 


\section{Introduction}

 It is well known, that certain galaxies host an active galactic nuclei (AGN) in their center. These astrophysical objects constitute the most luminous non-transient objects in the Universe \citep{Banados2018}. AGNs are also of interest in galaxy formation theory, as they are currently considered as a possibly important source of feedback (see e.g. \citealt{Silk1998}; \citealt{Sijacki2007}; \citealt{Booth2009}; \citealt{Fabian2012}; \citealt{Richardson2016}).  
A variety of different classes of AGN -- discovered independently in different wavelength regimes -- have been categorized over the past decades \citep[see ][for a review]{Padovani2017}. Among these classes, there is a general division based on the spectroscopic properties of the AGN in type I and type II dependent on, respectively, the presence or absence of broad emission lines 
(typically with ${\rm FWHM}\gg1000$ km/s). 
The so-called classical AGN Unification Model tries to explain the difference between type I and type II AGNs as due to presence of a dusty, circumnuclear torus and its orientation with respect to the observer line-of-sight \citep{Antonucci1993,Urry1995,Netzer2015}.  A distinct prediction of this model is that ionising radiation should be able to propagate to large distances only within an ``ionisation cone" around an axis perpendicular to the obscuring torus. In the low-redshift Universe, these ionisation cones have been observed and studied in numerous previous works \citep[for example][]{Schmitt1996,Jaffe2004,Afanasiev2007, Muller2011,Fischer2013, Durre2018}. At higher redshift the picture is less clear. 

To our knowledge, there are no direct constraints on the actual geometry of the AGN emission at high redshift. For instance, it is still not clear whether the obscuration is due to individual clump(s) along our line-of-sight rather than an actual torus or circumnuclear distributed clumps as suggested by the classical AGN unification model. A few studies that investigate the CGM of quasars at redshift $z\sim 2-3$ by looking at the line-of-sight absorption \citep{Prochaska2013,Johnson2015,Cai2017} find indeed indication of anisotropic radiation, which is in accordance with the classical unification theory. 

In order to test the validity of the classical AGN unification scheme, it is thus necessary to directly observe or trace the geometry of the AGN emission as it propagates on scales larger than the circumnuclear regions, e.g. in the circumgalactic medium (CGM) and intergalactic medium (IGM) outside of the AGN host-galaxy.
The recent detection of fluorescent Ly$\alpha$ emission produced by bright quasars at $z>2$ (e.g., \citealt{Cantalupo2012}; \citealt{Cantalupo2014}; \citealt{Martin2014}; \citealt{Hennawi2015}; see \citealt{Cantalupo2017} for a review) opened up a new window of opportunity to perform such a study. 
In particular, new observational techniques based on Integral-Field-Spectrographs with large field-of-view, such as Multi Unit Spectroscopic explorer (MUSE) (see e g., \citealt{Bacon2015} for a description),  have revealed that bright fluorescent Ly$\alpha$ Nebulae extending by more than 100 physical kpc are nearly ubiquitous around 
quasars (i.e., type I AGN) at $z>3$ (e.g., \citealt{Borisova2016}; \citealt{Fumagalli2016}; \citealt{North2017}; \citealt{Ginolfi2018}; \citealt{Battaia2019}; \citealt{Marino2019}). These observations provide three-dimensional maps that directly trace both the distribution of warm ($T\sim 10^4$ K), dense gas around quasars 
(see e.g., \citealt{Cantalupo2019} and \citealt{Pezzulli2019} for recent quantitative constraints) and, in the recombination radiation scenario for Ly$\alpha$ emission, the geometry of the quasar ionising radiation. 
In the assumption that the gas distribution on CGM and IGM scales around type I and type II AGNs is similar, any differences in the morphological properties of these nebulae
could thus reflect differences in the ``illumination" geometry. 

While type-I AGNs have been routinely targeted for fluorescent Ly$\alpha$ emission surveys or for other science purposes (providing a sample of more than 100 nebulae spanning a large range of redshifts from $z\sim2$ to $z\sim6$), type II AGNs have 
been rarely observed so far, mostly because they are more difficult to find than quasars at high redshift. This is due to their much fainter continuum magnitudes (due to obscuration along our line of sight) and the need of spectroscopic follow-ups
to confirm their AGN nature. This means that type II AGN catalogues are usually restricted to fields with deep multi-wavelength observations with spectroscopic follow-ups, e.g. the Chandra Deep Field \citep{Luo2017}, or serendipitously discovered. 
Here we take advantage of MUSE observations around 4 type-II AGNs, three of which have been observed as part of the MUSE--WIDE and MUSE--Deep surveys (\citealt{Urrutia2019}; \citealt{Bacon2017}) and one serendipitously discovered during a MUSE follow-up of a Ly$\alpha$ nebula detected in a narrow-band survey for fluorescent Ly$\alpha$ emission in a quasar field at $z\sim3$ (\citealt{Borisova2016b}; \citealt{Marino2018}; \citealt{Cantalupo2007}). We highlight the source Cdfs 15 (formally Cdf 202), which constitutes one of the first X-ray discovered Type II AGN at high redshifts \citep{Norman2002}. We stress that we did not include radio galaxies in our sample, as in the Seyfert classification scheme, Type II AGNs are always radio-quiet. 
Despite the limited sample, our study is one of the first attempts to compare the morphologies of nebulae around different types of AGN and provides new constraints on the properties of AGN emission at high redshift. 

The paper is organized as follows: In \autoref{sec_data} we describe the sample used in our analysis as well as the data processing and reduction that is applied. In \autoref{sec_analysis} we talk about the analysis method used to detect the extended Ly$\alpha$ and \heii\ emission. The results are presented in \autoref{sec_result} and discussed in \autoref{sec_disc}. Eventually, in \autoref{sec_conc}, we conclude this study. In the appendix, we present additional figures.
For calculations of cosmic distances in this study we use the $\Lambda$CDM cosmology and assume $\Omega_M = 0.3$, $\Omega_\Lambda=0.7$ and $h =70$ kms$^{-1}$Mpc$^{-1}$. Categorization of AGNs into type I (Seyfert 1) and type II (Seyfert 2) follows \cite{Osterbrock1981}, which is based on the width of the Balmer emission lines in the optical wavelength regime.


\begin{table*}
            \centering
            \caption[Sample Description]{Sample of Type II AGNs.}
        \label{tab:Sample_sum}
        \begin{tabular}{c c c c c c c c c c }
             \hline
            Number & AGN Name & Obj. ID$^{\text{a}}$ & R.A.  & decl. & $z_{\text{cat.}}^{\text{b}}$ &$z_{\text{Ly}\alpha}$ $^{\text{c}}$ &Exp. Time  & $L_{\rm X,c}^{d}$ & i$_\text{AB}^e$\\
              &  &  &(J2000)&(J2000) & & &[hr]  & [erg s$^{-1}$]&[mag]\\ \hline \hline
            1 & Bulb & -- & 04:22:01.5&--38:37:19.0 & --  &3.0984 & 20 &  --&$24.5$\\
            2 & UDF 09 & 00005 &03:32:39.7&--27:48:50.2& 3.072& 3.0687 &10 & $8.0\times10^{44}$ &$22.4$\\
            3 & Cdfs 04  & 05479&03:32:18.8&--27:51:35.5&3.661& 3.6620 &1& $2.8\times10^{44}$ &$24.5$\\
            4 & Cdfs 15 &06294&03:32:29.8&--27:51:05.9&3.710& 3.7027 &1 & $4.5\times10^{44}$& 25.7 \\
        \end{tabular}
        \vspace{0.3 cm}
    
        \raggedright
        
        { \footnotesize Notes. \\
        $^{\text{a}}$ Object ID taken from \cite{Guo2013} catalogue.\\
        $^{\text{b}}$ Galaxy redshifts from \cite{Inami2017} \\
        $^{\text{c}}$ Obtained from the flux-weighted wavelength centroid of the nebular Ly$\alpha$ emission.\\
        $^{\text{d}}$  Absorption-corrected intrinsic rest-frame (0.5 -- 7 keV) luminosity, from Chandra 7Ms catalogue \citep{Luo2017}, taking the intrinsic absorption column density $N_{\rm H}$ into account.\\
        $^{\text{e}}$ Apparent magnitude within the SDSS i-band filter using a circular aperture of 3$\arcsec$ diameter.
        }
    
\end{table*}

\section{Observational Data and Reduction}

\label{sec_data}
The observations were carried out using the MUSE instrument \citep{Bacon2010}, which is a multi-unit integral field spectrograph mounted on the ESO Very Large Telescope (VLT). 
MUSE provides a large field of view (1$^\prime\times$1$^\prime$) and, at the same time, a large number of spatial (about $300 \times 300$ pixels, each with a size of about $0.2\arcsec\times0.2\arcsec$) 
and spectral resolution elements  (about $3641$, covering $4750-9350$ \AA\ in the ``nominal" wavelength-range mode, which yields a wavelength layer scale of 1.25 \AA) in a single pointing. 
This gives us the possibility to study the morphology derived from the extended emission on large scales (up to about 450 physical kpc at $z\sim3$) at high spatial resolution (seeing limited at about $0.6$ arcsec or about 4.5 physical
kpc at $z\sim3$) without any constraints on source redshifts, provided that $z>2.95$ (for the nominal wavelength-range mode), and without narrow-band filter losses. 

\subsection{Sample Selection \& Observation}
Our sample consists of 4 type II AGNs, of which the Cdfs 04 and Cdfs 15 have been selected using the MUSE--WIDE survey \citep{Urrutia2019}, UDF 09 is found in the MUSE--Deep survey \citep{Bacon2017} and the Bulb AGN was taken from \cite{Marino2018}. 
The selection from the MUSE--WIDE survey was done by querying SIMBAD and Chandra-7 Ms X-ray survey \citep{Luo2017} for AGNs with spectroscopic redshifts between 3 and 4. We found an additional source (in UDF 02). However, we decided to remove it from the sample as it did not show any MUSE optical AGN lines and was much lower in X-ray luminosity by 2 orders of magnitude than the other sources.
The redshift range was set to $3 \lesssim z\lesssim 4$ to guarantee wavelength coverage of the Ly$\alpha$ emission line and to make sure that \heii\ $\lambda$1640 
emission line is not shifted too much into the wavelength range affected by telluric lines ($\geq 6500$~\AA). The observations are part of various observation programs, mainly of MUSE GTO observations. As a consequence, we have different exposure times for different sources, ranging from 1 h up to 20 h. The sample properties are summarized in \autoref{tab:Sample_sum}. \autoref{fig:White_Light} shows the white light image as well as the integrated spectrum of these sources.
As a control Type I sample, we adopt the one used and described in \cite{Borisova2016}, which consists of 17 radio--quiet quasars and 2 radio--loud quasars. They used the \cite{Veron2010} catalog to select quasars, which is another terminology for type I AGN. The exposure time for these observations was 1 h.


\subsection{Data Reduction}
The data reduction follows the steps as described in \cite{Borisova2016} and \cite{Marino2018}.
In particular, we first reduce the cubes with the standard ESO MUSE Data Reduction Software (v1.6, \cite{Weilbacher2015}) with the default (recommended) parameters and without performing sky-subtraction.
Then, we run post-processing optimisation of the flat-fielding and sky subtraction using the \texttt{CubeFix} and  \texttt{CubeSharp} tools, part of the \texttt{CubExtractor} (hereafter called \texttt{CubEx}, v1.8) package (Cantalupo in prep., see \citealt{Cantalupo2019} for a description).
Finally, all individual exposures have been combined with an average-sigma clipping method using the  \texttt{CubeCombine} tool (also part of  \texttt{CubEx}). 
These steps have been also applied to the datacubes of the MUSE--WIDE survey. The field associated with the UDF 09 source has been instead reduced as described in \cite{Bacon2017} using a flat-fielding optimisation similar to  \texttt{CubeFix}. 


\section{Analysis}
\label{sec_analysis}
 \subsection{Post--processing of the MUSE cubes}   
We exploit the MUSE integral field unit (IFU) capabilities to detect and study extended Ly$\alpha$ and \heii\ emission around the type II AGNs. 
While quasar fields are usually post-processed with quasar-PSF subtraction algorithms (see CubePSFSub routine described in \citealt{Cantalupo2019}),
we do not perform this step in our analysis of type II AGNs. Indeed, because the  central nucleus is obscured along our line of sight for these systems,
no PSF subtraction is needed to reveal the central parts of extended line emission associated with the nebulae (or the AGN host galaxy). \autoref{fig:SBProf_comp} in the appendix shows indeed that the PSF does not contribute significantly to the nebula in the case of the bulb nebula. In this plot, the orange line indicates the surface brightness (SB) profile of the PSF while the blue line is the circularly averaged SB profile of the bulb nebula.  
Continuum subtraction is performed with the median-filtering routine \texttt{CubeBKGSub}, as described in several previous works \citep[e.g.][]{Cantalupo2019}. In particular,  a  median-filtering bin-width of 40 pixel is used and the result is additionally smoothed across four adjacent bins. 
Before median-filtering, the spectral regions with expected emission lines are masked to avoid over-subtraction. 
Finally, we divide the cube into subcubes by selecting only a range of slices where we expect the Ly$\alpha$ or the \heii\ emission line (around 100--200 slices, i.e. $\Delta \lambda = 125-250$ \AA).

\begin{figure*}
    \centering
    \includegraphics[width = \textwidth]{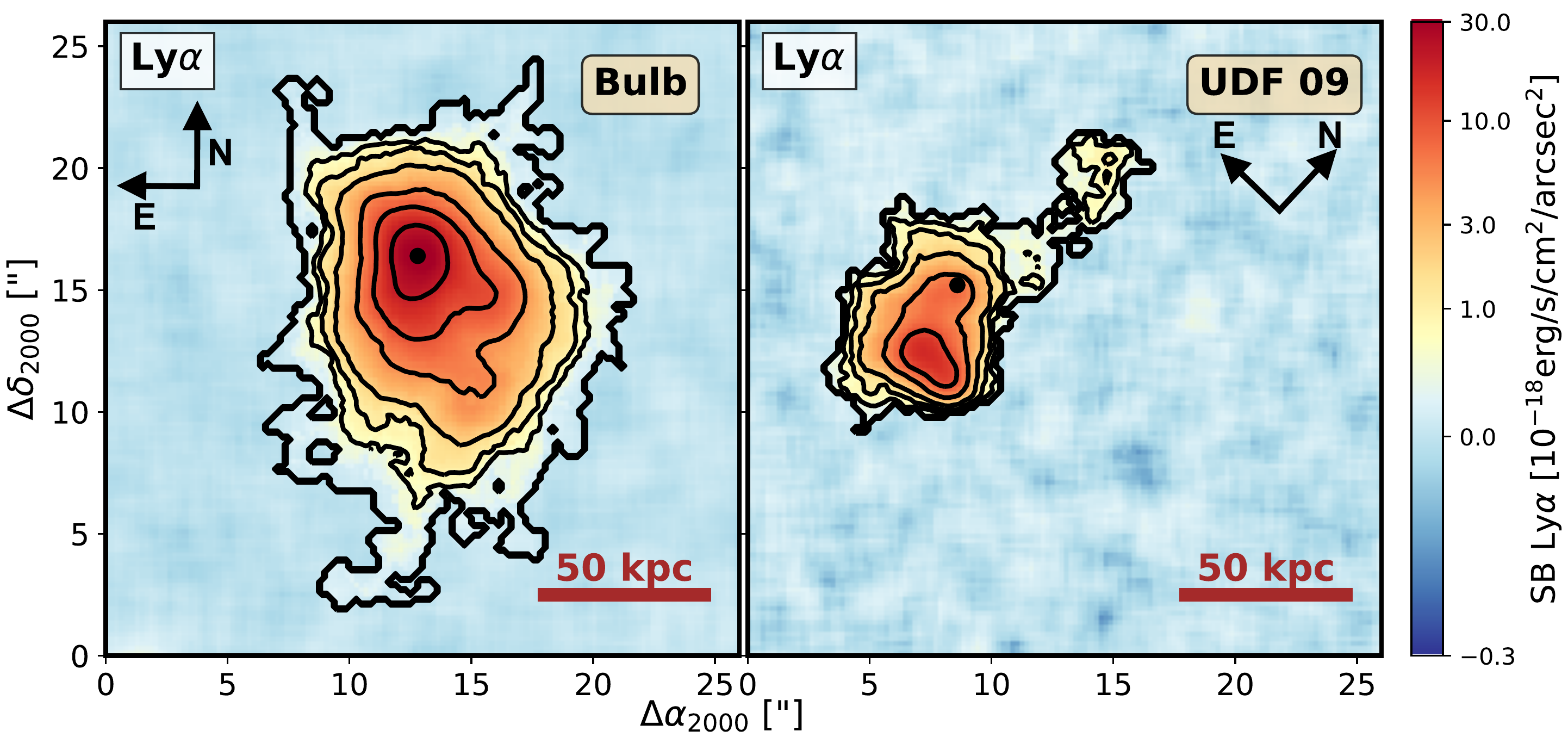}
    \caption{Optimally extracted SB maps of the detected Ly$\alpha$ emission for the Bulb (left) and UDF 09 (right) source after continuum subtraction (see text for details). 
    The thin contours show increasing S/N levels corresponding to SB = 0.5, 1, 2, 5, 10, 20 ($10^{-18}$ erg/s/cm$^2$/arcsec$^2$). The thick contour indicates the S/N = 2 isophote. Outside this contour, a single MUSE wavelength layer close to the central wavelength of the detected emission is shown. The black dot indicates the AGN position. The dark--red bar indicates a physical scale of 50 kpc. Both sources show asymmetric spatial morphologies. The Bulb nebula is more extend towards the south--west. Note that the brightest Ly$\alpha$ regions in the UDF 09 field is not coincident with the AGN position. }
    \label{fig:SBmaps}
\end{figure*}
\begin{figure*}
    \centering
    \includegraphics[width = \textwidth]{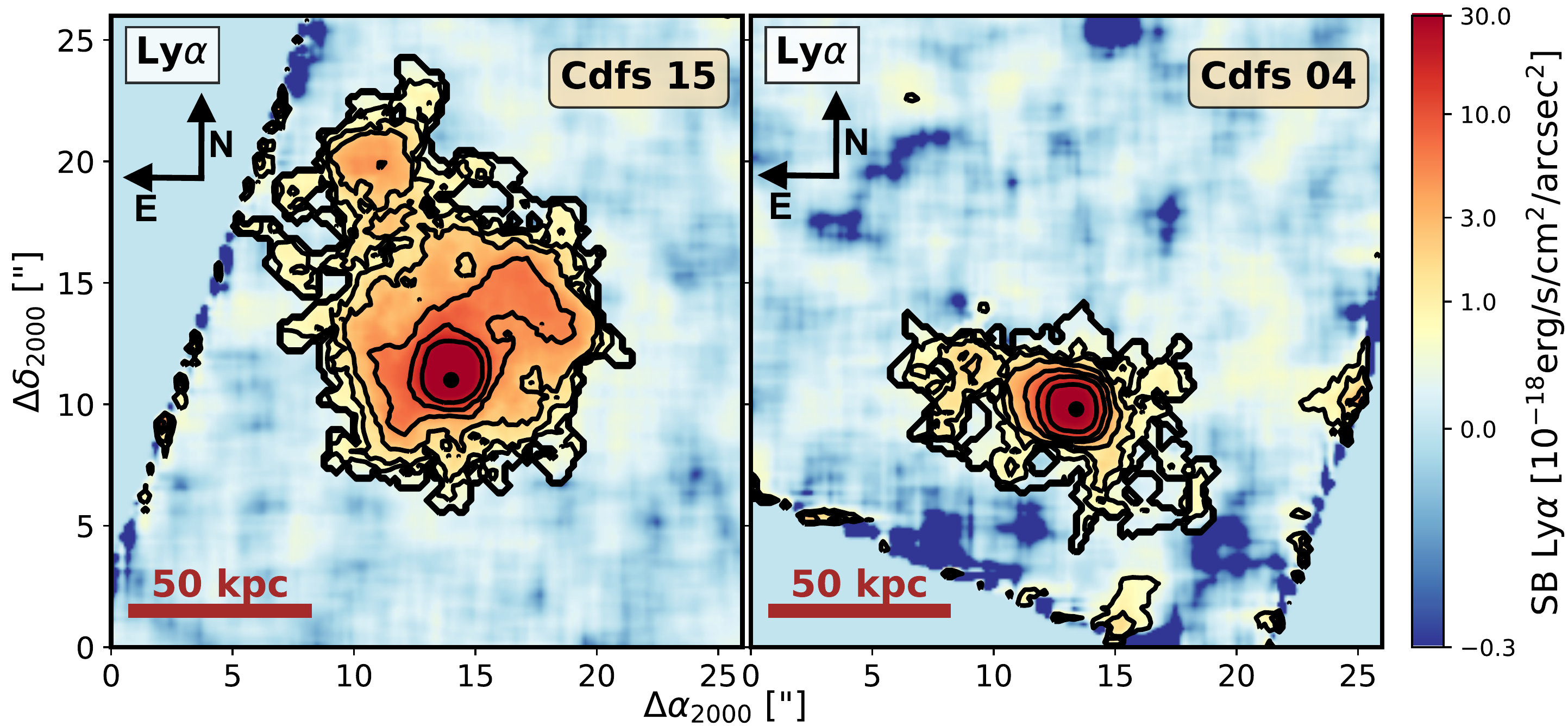}
    \caption{Optimally extracted SB maps of the detected Ly$\alpha$ emission for the Cdfs 15 (left) and Cdfs 04 (right) source obtained as in \autoref{fig:SBmaps}. Note that these two fields have been observed only for 1 hour of total exposure time while the Bulb and the UDF 09 field have been observed for 20 hours and 10 hours, respectively. In addition, the sources are at higher redshifts than the other two sources. The Cdfs 15 shows slightly asymmetric spatial morphology towards the north--east. The Cdfs 04 Ly$\alpha$ nebula is the faintest in our sample and it is clearly elongated.}
    \label{fig:SBmaps_2}
\end{figure*}
\begin{figure*}
    \centering
    \includegraphics[width = \textwidth]{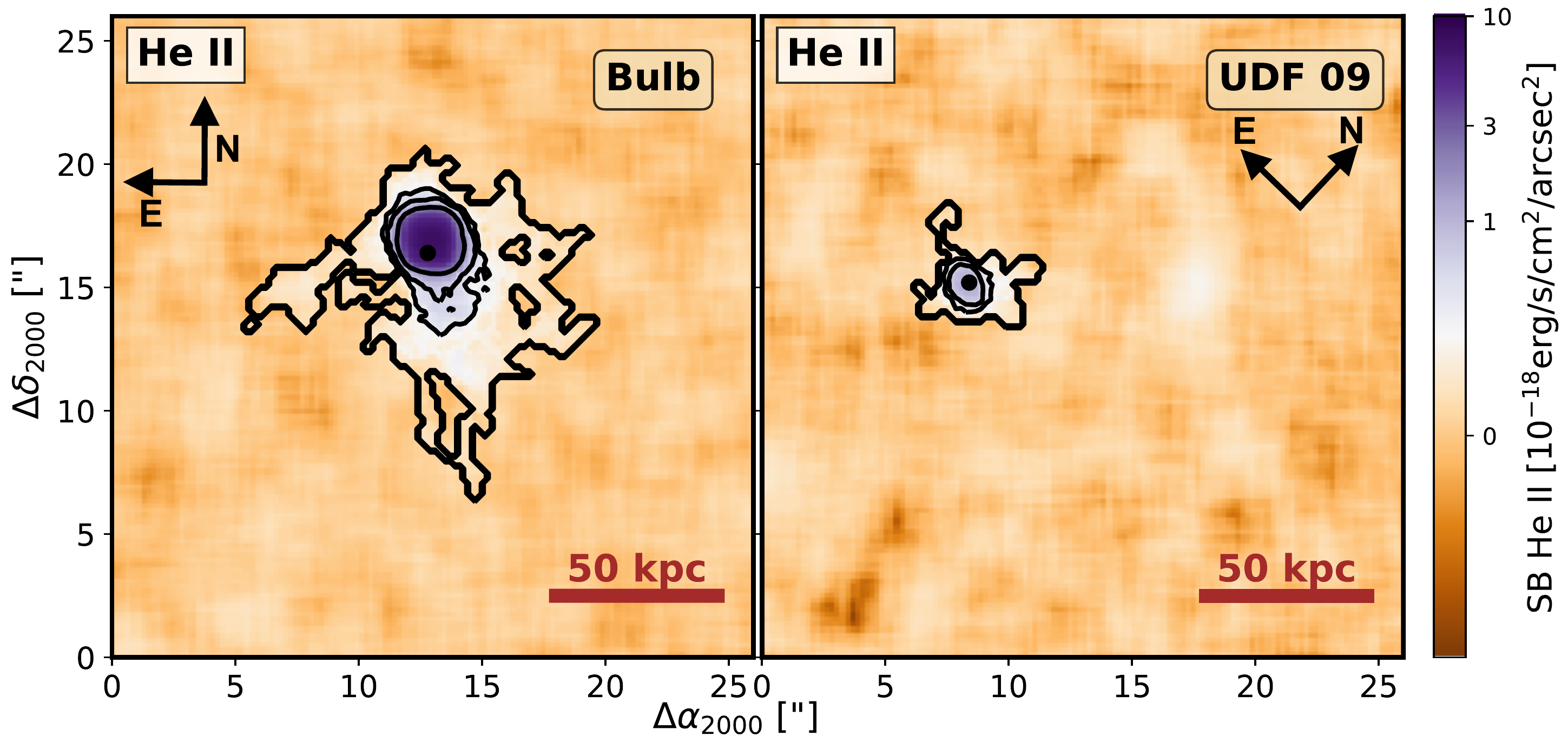}
    \caption{Optimally extracted SB maps of the detected \heii\ $\lambda$1640 extended emission for the Bulb (left) and compact emission for the UDF 09 (right) sources obtained as in \autoref{fig:SBmaps}.
    The levels of the thin contours correspond to SB = 0.15, 0.5, 2 ($10^{-18}$ erg/s/cm$^2$/arcsec$^2$).
    The Bulb shows extended \heii\ $\lambda$1640 emission out to 50 projected physical kpc from the AGN position. The emission is clearly more extended towards the south--west. The UDF 09 nebula \heii\ $\lambda$1640 emission is much less extend. We note that the region with the brightest Ly$\alpha$ emission (in the UDF 09 source) does not show any detectable \heii\ emission at the current depth. 
    }
    \label{fig:SB_images_HeII}
\end{figure*}
\subsection{Detection of Ly\texorpdfstring{$\alpha$}{TEXT} and \heii\ Extended Emission}
    The 3D detection and continuum subtraction is performed using the \texttt{CubEx} routine 
    which extracts contiguous regions in the datacube above a given signal-to-noise ratio (S/N). 
    The implementation is based  on a connected-labelling-component algorithm extended to three dimensions. That way, voxels above a chosen threshold are connected. It is counted as a detection if the number of connected voxels exceeds a certain threshold.
    The following input parameters are used for the source extraction with \texttt{CubEx}: (i) a Gaussian spatial filter of radius $0.4\arcsec$ is chosen (we did not smooth in the wavelength direction); (ii) S/N threshold is set to 2 after spatial smoothing (corresponding to a surface brightness (SB) $\sim10^{-18}$ erg s$^{-1}$ cm$^{-2}$ arcsec$^{-2}$ using an aperture of 1 arcsec$^{2}$, a single wavelength layer of
    1.25 \AA, and an exposure time of 1 h); (iii) we mask the bright continuum sources and regions associated with large sky residual noise in the MUSE cube; (iv) we rescale the pipeline--propagated variance to match the actual (spatial) variance measured across each layer of the datacube; (v) we set the minimum number of connected voxels to $N \ge 1000$. 
    This procedure yields a 3D segmentation map as well as a catalogue of all line emitting sources detected in the MUSE cube.
    
    We detect extended Ly$\alpha$ emission for all four type II AGN sources. The nebulae have maximum projected radial sizes ranging from 50 to 100 physical kpc (kpc) (see \autoref{tab:Neb_prop}), as revealed from the optimally extracted images shown in  \autoref{fig:SBmaps} and \autoref{fig:SBmaps_2}, obtained from integrating the flux along the wavelength direction across the spectral layers identified in the 3D segmentation map as commonly done in the literature (e.g., \citealt{Borisova2016}). 
    The 2D projection of the segmentation mask is indicated by the thick contour line. Outside this contour, we show the SB of a single layer to give an idea of the noise associated with the faint and more diffuse parts of the nebulae. 
    The wavelength of the layer has been chosen to be close to the central wavelength of the detected emission \citep[find a more detailed description in][]{Borisova2016, Cantalupo2019}. It is important to note that these images differ from (pseudo) narrow band images (see \autoref{fig:NB_images} and \autoref{fig:NB_images_2} in the appendix) because they are optimised in S/N by tailoring the spectral width for every individual pixel, avoiding filter losses.
    With this technique, we can get deep sensitivity levels and a high flux dynamic range for MUSE cubes with exposure times of 1 h that are 
    comparable to 20 h observations with NB filters (40 \AA\ filter) on an 8--meter telescope \citep{Cantalupo2012}.  In order to derive the SB radial profiles
    we use instead pseudo-NB images integrated along a fixed number of layers in order to recover all the possible flux also for the fainter and more diffuse emission
    in the external regions (the layer width is indicated in \autoref{tab:Neb_prop} in the last columns and is determined in a way to include all the emission of the line with S/N>2.). Consequently, the pseudo NB images have a spatially uniform noise level  as opposed to the optimally extracted ones.
    
    In addition to Ly$\alpha$ emission, we also searched for extended \heii\ $\lambda$1640 emission in the proximity of the AGN in our deepest dataset (Bulb and UDF 09). 
    The Bulb shows extended \heii\ emission out to a maximum radial distance of 50 pkpc. The UDF 09 source shows compact emission with only faint extended substructures out to 25--30 pkpc from the center of the AGN. This difference in detected sizes is not necessarily physical, considering the different depth of the observations. The optimally extracted SB images can be seen in \autoref{fig:SB_images_HeII}. The Ly$\alpha$ and \heii\ nebulae properties are summarized in \autoref{tab:Neb_prop}. 
 
\begin{table}
            \centering
            \caption[Sample Description]{Measured Nebulae Properties}
        \label{tab:Neb_prop}
        \begin{tabular}{c c c c c c}
             \hline
            Name & $z_{\text{Ly}\alpha}^{\text{a}}$ & Line & $r_\text{max} ^{\text{b}}$ & Flux$^{\text{c}}$ &$\Delta\lambda^{\text{d}}$\\
            &&&[pkpc]&[erg s$^{-1}$ cm$^{-2}$] & [\AA]\\\hline \hline
             \multirow{2}{*}{Bulb}& \multirow{2}{*}{3.0984} & Ly$\alpha$ & 75 &$6.6\times10^{-16}$&58.8\\
             &&\heii\ &50&$4.0\times10^{-17}$&40\\\hline
             \multirow{2}{*}{UDF 09} & \multirow{2}{*}{3.0687} &Ly$\alpha$ & 62 &1.7$\times10^{-16}$& 17.5\\
             &&\heii\ &40&$6.6\times10^{-18}$&22.5\\\hline
             \multirow{2}{*}{Cdfs 04}  &\multirow{2}{*}{3.6620}&Ly$\alpha$&48 &$2.1\times10^{-16}$&39\\
             &&\heii\ &--&--&--\\\hline
             \multirow{2}{*}{Cdfs 15} &\multirow{2}{*}{3.7027}&Ly$\alpha$&92 &$5.7\times10^{-16}$&48.8\\
             &&\heii\ &--&--&--\\
        \end{tabular}
        \vspace{0.3 cm}
        \raggedright
        
        { \footnotesize Notes.\\$^{\text{a}}$ Obtained from the flux-weighted wavelength centroid of the nebular Ly$\alpha$ emission.\\
        $^{\text{b}}$ Maximal extended radius from the AGN position using a 2D projection of the 3D segmentation map (see text for details).\\
        $^{\text{c}}$ These fluxes are extracted over different radii/segmentation masks. \\
        $^{\text{d}}$ Maximal depth in the spectral dimension of the 3D segmentation map (see text for details).}
    
\end{table}


\section{Results}
 \label{sec_result}   
As shown in the previous section, all 4 type II sources have a detected Ly$\alpha$ nebula. Here, we describe in detail their
morphological properties.

\subsection{Visual Inspection of Nebulae Morphology}
The extended Ly$\alpha$ emission around the four sources shows a variety of shapes and sizes, as can be seen in \autoref{fig:SBmaps} and \autoref{fig:SBmaps_2}. 
The Bulb Ly$\alpha$ nebula is extended towards the south--west. Also the Bulb's \heii\ emission extends towards the south--west. Similarly, the Cdfs 15 nebula is extended more in one direction -- towards the north--east. We point out that the source is surrounded by at least two galaxies within the Lya halo, particularly in the north-east. This second galaxy is situated at the same redshift as the AGN. The large SB and extension of the part of the nebula surrounding this companion galaxy is however incompatible with ionisation from the companion galaxy itself (Ly$\alpha$ haloes commonly observed around galaxies are indeed at least one order of magnitude fainter and smaller than this nebula), leaving the AGN as the most plausible source of ionisation.
The Cdfs 04 nebula is the smallest in our sample, showing a maximal diameter of approximately 80 pkpc. While the innermost part is slightly asymmetric towards the east, the parts further out show a clear elongated structure from east to west.  
The UDF 09 is more peculiar: the strongest Ly$\alpha$ emission is not coming from the central region of the AGN, but rather from a region towards the south--east. In addition, there is an asymmetry towards the north--west. 
The measured type II AGN nebulae properties are listed in \autoref{tab:Neb_prop}. 

The morphology of the type II AGNs show hints to be qualitative different than what typically seen around the 19 type-I AGNs in the \cite{Borisova2016} sample. Most of the type I nebulae are indeed quite spatially symmetric and circular.   
in particular, only two sources in the type I sample (\#1 and \#3) show clearly extended nebulae with asymmetric (and filamentary) patterns at high S/N level. We are aware that a visual inspection is subjective. We will continue the analysis using a quantitative measure of asymmetry.  

\subsection{Quantifying the Spatial Asymmetry of Ly\texorpdfstring{$\alpha$}{TEXT} Nebulae}

For a more quantitative comparison between the type II and type I samples, we perform a quantitative analysis of the Ly$\alpha$ extended emission's spatial morphology in terms of geometrical asymmetry and structure.
In particular, we use two methods to quantify the spatial morphology of the extended Ly$\alpha$ emission, both based on the optimally extracted images. 

As a first method, we use a modified version of the asymmetry quantificator described in \cite{Stoughton2002} and recently used also by \cite{Battaia2019}.

\begin{enumerate}
    \item we first determine the AGN position  $(x_\text{AGN}, y_\text{AGN})$ from the continuum, white-light image (which corresponds to the position of the AGN)
    and we use this position for the calculation of the second--order moments of the pixel distribution 
    (note that this is different from the procedure of \cite{Battaia2019}, in which the flux--weighted centroid of the Ly$\alpha$ emission is used);
    \item we calculate the second--order moments with respect to the AGN position for all spatial pixels within the area associated to the nebula (
    defined as the region above the 2$\sigma$ isophote in the optimally-extracted images),
    without applying any flux-weighting:
\begin{equation}
\begin{split}
    M_{xx} &:= \left<\frac{(x - x_\text{AGN})^2}{r^2}\right>,
    M_{yy} := \left<\frac{(y - y_\text{AGN})^2}{r^2}\right>\\
    M_{xy} &:= \left<\frac{(x - x_\text{AGN})(y - y_\text{AGN})}{r^2}\right>
\end{split}
\end{equation}
where $r$ is the two-dimensional distance between the point $(x,y)$ and the AGN position. 
   \item
   finally, we quantify the asymmetry of the nebulae with the dimensionless parameter $\alpha$, which is defined by the following ratio of second--order moments (for a more detailed derivation see \citealt{Battaia2019}, section 4.1.1):
\begin{equation}
    \alpha = \frac{1-\sqrt{(M_{xx}-M_{yy})^2+(2M_{xy})^2}}{1+\sqrt{(M_{xx}-M_{yy})^2+(2M_{xy})^2}}
\end{equation}
\end{enumerate}
\begin{figure}
    \centering
    \includegraphics[width = 0.9\columnwidth]{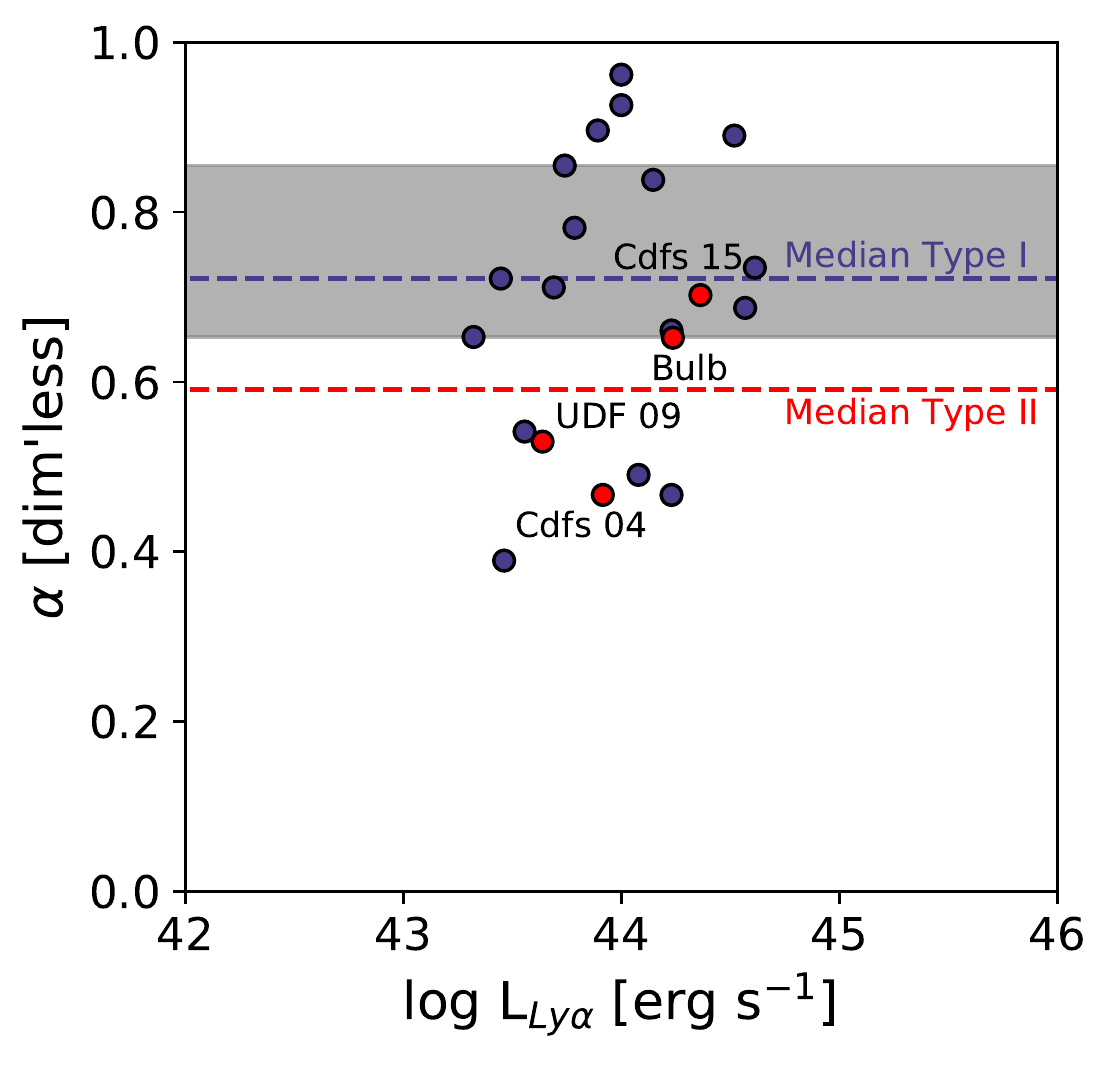}
    \caption{The asymmetry described by the parameter $\alpha$ based on the routine described in \protect\cite{Battaia2019} versus the Ly$\alpha$ nebula luminosity. Blue points indicate the type I sample (from \protect\cite{Borisova2016}) and red points indicate the type II AGN sample. The shaded area indicate the 25 and 75 percentile of the type I sample $\alpha$ distribution. The type II AGN's median is below the 25 percentile of the type I population indicating higher asymmetry.}
    \label{fig:method_Battaia}
\end{figure}
\begin{figure*}
    \centering
    \includegraphics[width = 0.9\textwidth]{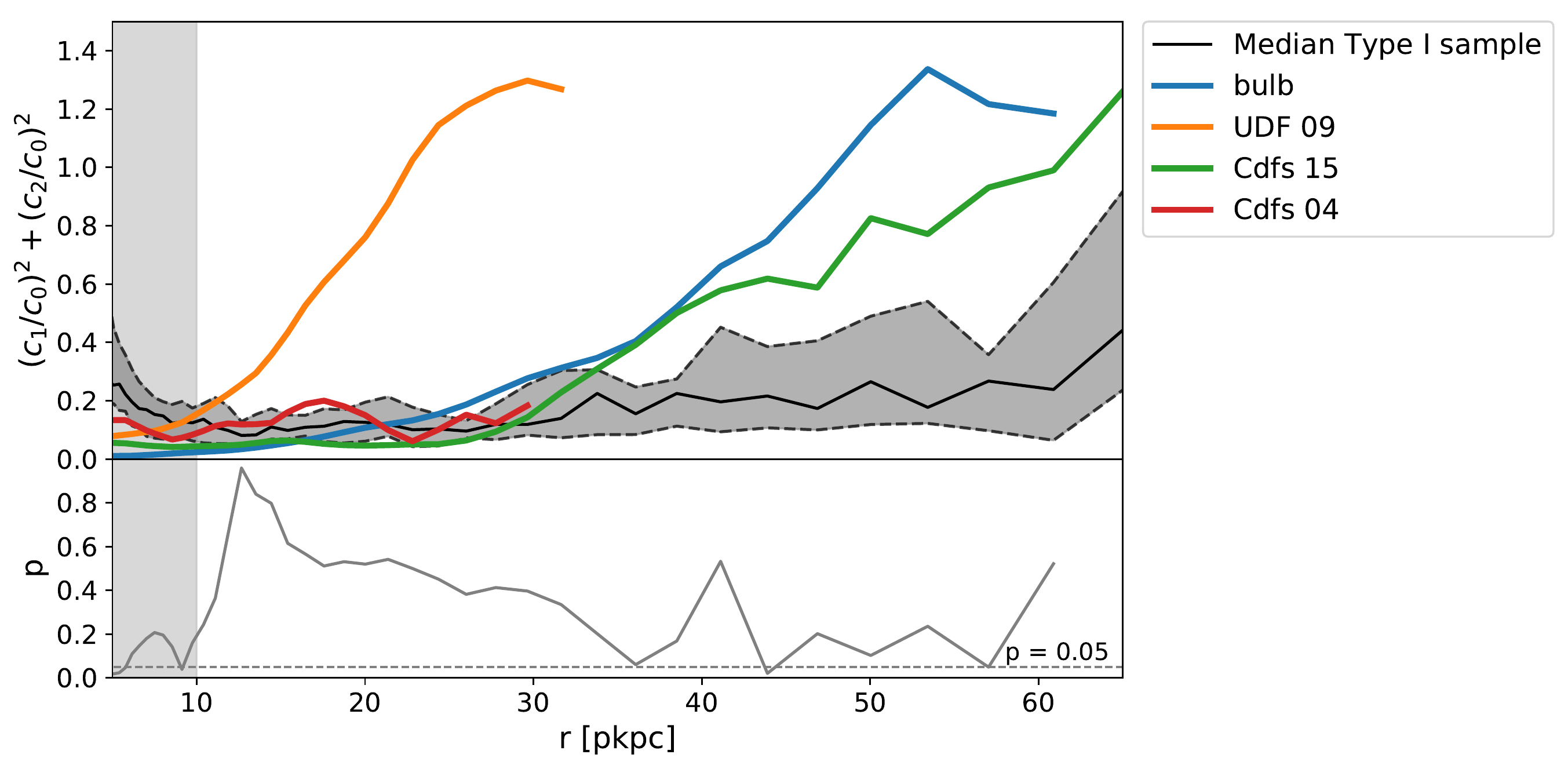}
    \caption{Fourier analysis performed on the four type II AGN sources (colored lines) as well as the type I AGN sample from \protect\cite{Borisova2016} (median indicated by black line). The dashed line and the shadowed region indicate the 25 and 75 percentile around the type I AGN median. We used a logarithmic binning for the radial distances $r$. The deviation from spherical symmetry is captured by the ratio of the Fourier coefficients $(c_1/c_0)^2 + (c_2/c_0)^2$. A larger value indicates that higher modes become dominant over the spherical mode. This means that the source shows a higher asymmetry. The bottom panel shows the result from the t-test where for every radial distance $r$ we test whether the values from the type I and type II follow the same distribution. With increasing radial distance we observe, that the type II AGNs are more asymmetric than the type I AGNs, also captured by the lower $p$--values at larger distances. The vertical shaded area indicates the extend where PSF convolution effects might still be relevant. }
    \label{fig:Fourier_decomp_result}
\end{figure*}

A value of $\alpha = 1$ indicates very circular spatial morphology, while $\alpha<1$ implies a more asymmetric shape. The same procedure is applied both for the type II sample and for the optimally extracted images in the type I sample
of \cite{Borisova2016}.
Our modifications to the scheme previously used in \cite{Battaia2019} allow us to better characterise asymmetries on large scales 
\emph{with respect to the sources of interest in this study, i.e. the AGNs}. The flux-weighted moments are indeed dominated by the
central regions, that are much brighter than the external parts, and the nebulae flux-weighted centroids do not necessarily coincide the AGN positions (as it is clear, e.g., in our UDF 09 source). 

\autoref{fig:method_Battaia} presents the asymmetries $\alpha$ calculated with the method outlined above for both our sample and the type I sample from \cite{Borisova2016}. In particular, we show
the asymmetry parameter in function of the Ly$\alpha$ nebula luminosity within the S/N~=~2 contour. 
The median of both samples in terms of the $\alpha$ paramter, as well as the 25 and 75 percentile of the type I AGN sample, are indicated. 
We observe that the type II AGNs typically show values of $\alpha$ lower than the type I population indicating higher levels of asymmetry. In particular, 
the median of the type II sample lies below the 25 percentile range of the type I AGN sample. 
A Welsch t-test returns a  p-value of $p = 0.097$ indicating a 10\% chance that the type I and type II population follow the same $\alpha$ distribution. 
It is interesting to note that type II and type I AGN nebulae have similar luminosities. 
This could suggest that what we see is not a second-order effect, related to differences in luminosities, but more likely a genuinely geometrical effect.

The second method we apply to quantify the spatial asymmetry of the Ly$\alpha$ nebulae is based on  Fourier decomposition of the spatial distribution of the nebulae SB.
In particular, we compute the nebulae SB in function of radial distance ($r$) and azimuthal angle ($\theta$),
and convert these into Fourier coefficients
as described in detail below.
In addition to being completely independent from the previous method, 
the advantage of this approach is the possibility to study the asymmetry as a function of distance from the AGN. 
Our procedure starts by calculating the surface brightness in radial and azimuthal-angle bins:
$$
    \text{SB}_{\text{Ly}\alpha}(x,y) \mapsto \text{SB}_{\text{Ly}\alpha}(r,\theta),
$$
where $r$ is the two-dimensional distance from the AGN and $\theta$ the azimuthal angle. 
Then, we decompose it using the Fourier coefficients defined by:
\begin{equation}
    \text{SB}_{\text{Ly}\alpha}(r,\theta) = \sum_{k=0}^\infty \big[a_k(r)\cdot\cos(k\theta) + b_k(r)\cdot \sin(k\theta)\big]
\end{equation}
where the Fourier coefficients $a_k(r)$ and $b_k(r)$, which account for the directional asymmetry for a given radial distance $r$, are defined by:
\begin{align*}
    a_k(r) &= \frac{1}{2\pi}\int_0^{2\pi}\text{SB}_{\text{Ly}\alpha}(r,\theta)\cdot\cos(k\theta)d\theta\\
    b_k(r) &= \frac{1}{2\pi}\int_0^{2\pi}\text{SB}_{\text{Ly}\alpha}(r,\theta)\cdot\sin(k\theta)d\theta
\end{align*}

Note that  $a_0(r)$ corresponds to the circularly average brightness profile. 
For a perfectly circularly symmetric SB map, the $0^{th}$ coefficient will be the only non-vanishing term, while for an centered elliptically shaped SB map, also the $2^{\text{nd}}$ mode will be significant, while for an off-centered ellipse, also the $1^{\rm st}$ mode will be significant.
In our analysis, we are not interested in the directional asymmetry of the extended Ly$\alpha$ nebulae. 
Rather, we want for every radial distance $r$ to have a measure of how much the symmetry diverges from a perfectly circular one. 
We mitigate other possible effects due to asymmetries in a particular direction, by combining the $a_k$ and $b_k$ coefficients as follows:
$$
    c_k(r) = \sqrt{a_k(r)^2 + b_k(r)^2}
$$
The significance of a certain modal asymmetry at a radial distance $r$ with respect to the circular mode is captured by the ratio $c_k(r)/c_0(r)$. We set $r=0$ at the center of the AGN (see black dot in the SB maps).

 In \autoref{fig:Fourier_decomp_result}, we show our results expressed in terms of the ratio $(c_1/c_0)^2 + (c_2/c_0)^2$
 that captures the strength of the first and second mode with respect to the circular mode. 
 We limit here our analysis to the first two non-circular modes because we have found that higher Fourier coefficients are most of the times smaller than these.
 Note that we perform our analysis up to a maximum radius where the circular mode is equal to the noise level to avoid spurious signals. 
 
The coloured lines indicate the type II AGN sources, while the black line is the median (for each $r$) of the type I AGN sample. The dashed line and the shadowed region indicate the 25 and 75 percentile around the type I AGN median.
The innermost 5 pixel region is masked because of possible PSF subtraction residuals for the type I cases. 
It is already clear from visual inspection that 3 out of 4 of the type II sources deviate towards higher values at increasing radial distance 
(see also \autoref{fig:NB_images} and \autoref{fig:NB_images_2} for an indication of where these radial distances are located in the images). 
This result indicates that 3 over the 4 nebulae around type II AGNs diverge from spherical symmetry more than their type I counterparts.  
This is quantified through a t-test at each $r$ (see bottom panel of \autoref{fig:Fourier_decomp_result}). For a radial distance beyond 25 kpc, we see that the Bulb and the Cdfs 15 nebulae are statistically more asymmetric, with p-values reaching values as low as 0.05. Note that our method is also able to catch very clearly the peculiar behavior of the UDF 09 source. The asymmetry of this nebula, prominent also at small radii, is driven by the fact that the brightest emission of Ly$\alpha$ is not coming from the center of the AGN but rather from a region towards the south--east. 
Instead, the Cdfs 04 source radial asymmetry appears more similar to the type I AGN cases in this method. This is because the central region of this source is very symmetric ($r<10$ pkpc) and at larger distances, the emission is very faint. 

\begin{figure}
    \centering
    \includegraphics[width = \columnwidth]{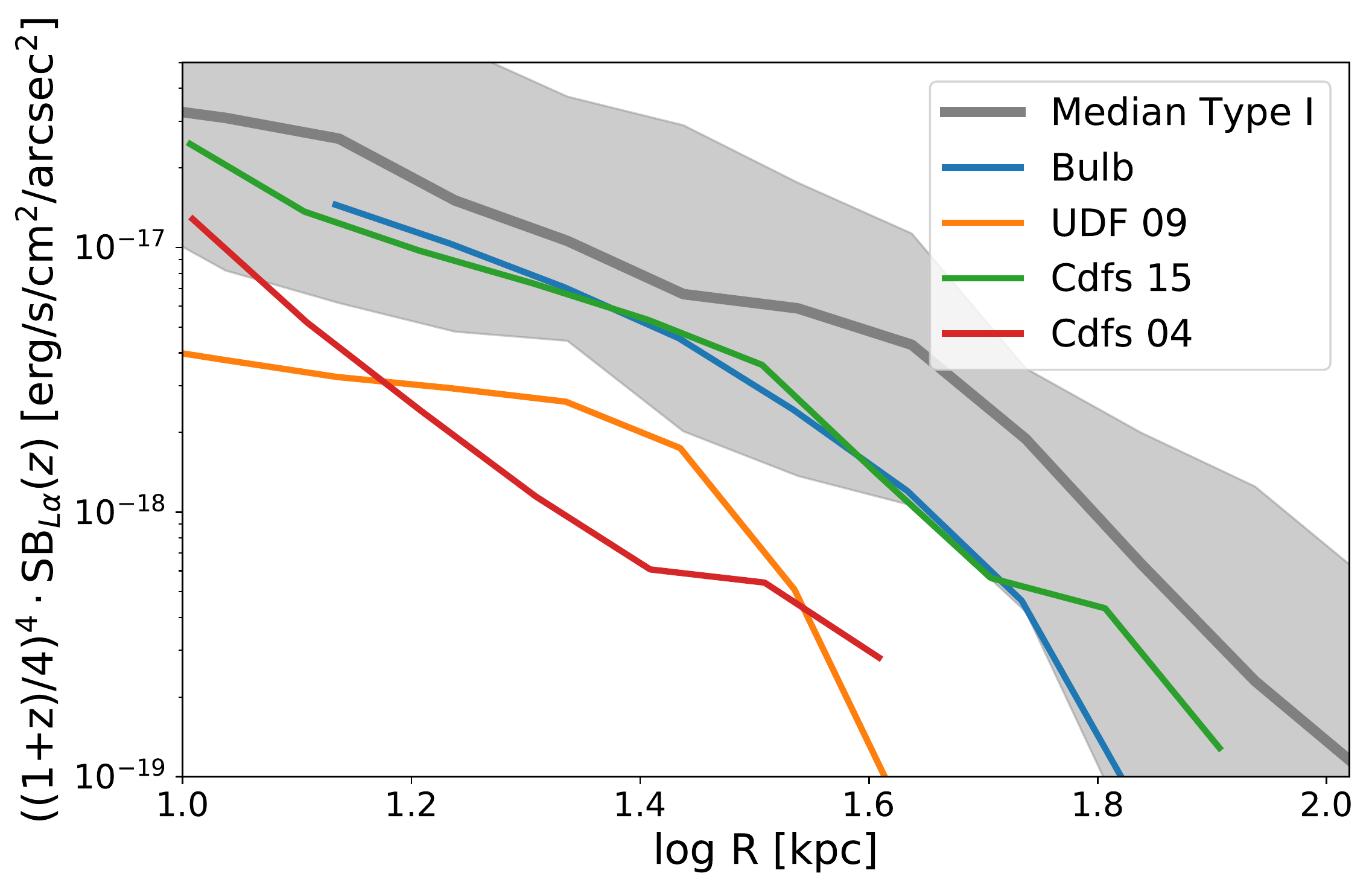}
    \caption{Circularly-averaged surface brightness profiles of the Ly$\alpha$ nebula for the type I and type II AGN sample. 
    The colored lines show the surface brightness profiles for the type II AGNs rescaled at $z=3$ by correcting for the redshift-dimming factor $((1+z)/4)^4$.
    The median profile of the type I AGN sources (values taken from \citealt{Marino2019}) has been also rescaled to redshift $z=3$ and is indicated by the solid grey line. The grey area represents the 10$^\text{th}$ and 90$^{\text{th}}$ percentiles of the type I distribution. While the circulary-averaged SB profiles of the type II AGN nebulae are typically fainter than their type I counterpart, they are still within the 10$^\text{th}$  percentile of the type I AGN SB profiles. Moreover, as discussed in this study, type II AGN nebulae are typically less circular than type I nebulae. This will likely contribute to produce fainter circularly-averaged SB profiles.}
    \label{fig:SBProfiles}
\end{figure}

\subsection{Surface Brightness Profiles}

Independent of our analysis above and for consistency with previous studies, we report here the circularly averaged surface brightness profiles in \autoref{fig:SBProfiles} for our sources (coloured lines) using the same method of \cite{Borisova2016} (the type I SB profiles are indicated by the grey line and shaded area).  
In particular, we use the median profile of the type I AGN sample from \cite{Borisova2016} as tabulated in \cite{Marino2019} and rescaled to redshift $z = 3$. 
The Bulb and Cdfs 15 profile lie within the 10$^{\text{th}}$ percentile of the type I median profile. The UDF 09 and Cdfs 04 nebulae are below the 10$^{\text{th}}$ percentile.  The individual surface brightness profiles of our type II AGN sample are also shown in the appendix in \autoref{fig:SB_profiles}. 
We fitted a power and an exponential law to the surface brightness profiles. The results of the fitting procedure are listed in \autoref{tab:Result_Fitting}. 
The Bulb nebula is better fitted by an exponential law, while the two Cdfs sources are better fit by a power law. In general, we see when comparing those with the profiles from \cite{Borisova2016} (figure 4 therein) that despite of the different morphology and sizes of the sources, the SB profiles look rather similar. Higher asymmetry will likely contribute to produce fainter circularly-averaged SB profiles.
Including the fact that the type II AGN nebulae are less circularly symmetric, as showed above, this suggests that, overall, the local SB values among the two samples are similar. 


\subsection{\heii\ Emission and Line Ratios}

Because Ly$\alpha$ is a resonant line, radiative transfer effects could play a role washing out possible asymmetries in the ``illumination" geometry,
especially in the central regions, limiting our possibility to address our motivating question in the inner parts of the nebulae.
For this reason, we searched for non-resonant lines such as \heii\ $\lambda$1640. Because of the faintness of this line with respect to Ly$\alpha$ (see e.g. \citealt{Cantalupo2019}, where the ratio is around 0.03$-$0.08), we do not expect clear detections in the shallower fields
such as the Cdfs sources. Indeed, we have been only able to detect \heii\ $\lambda$1640 emission for the Bulb and UDF 09 sources for which we have at least 10 hour exposure time. 

In particular, the Bulb nebula shows faint and extended \heii\ emission (see optimally extracted surface brighness maps in \autoref{fig:SB_images_HeII}) up to a distance of 50 kpc from the AGN (indicated by the black dot). The UDF 09 source shows mostly compact emission with some indications of 
an extended emission region by a few kpc around the AGN. Interestingly, the \heii\ $\lambda$1640 emission is also asymmetric, as in the Ly$\alpha$ emission case, but rather circular in the central regions. 
\begin{figure}
    \centering
    \includegraphics[width = \columnwidth]{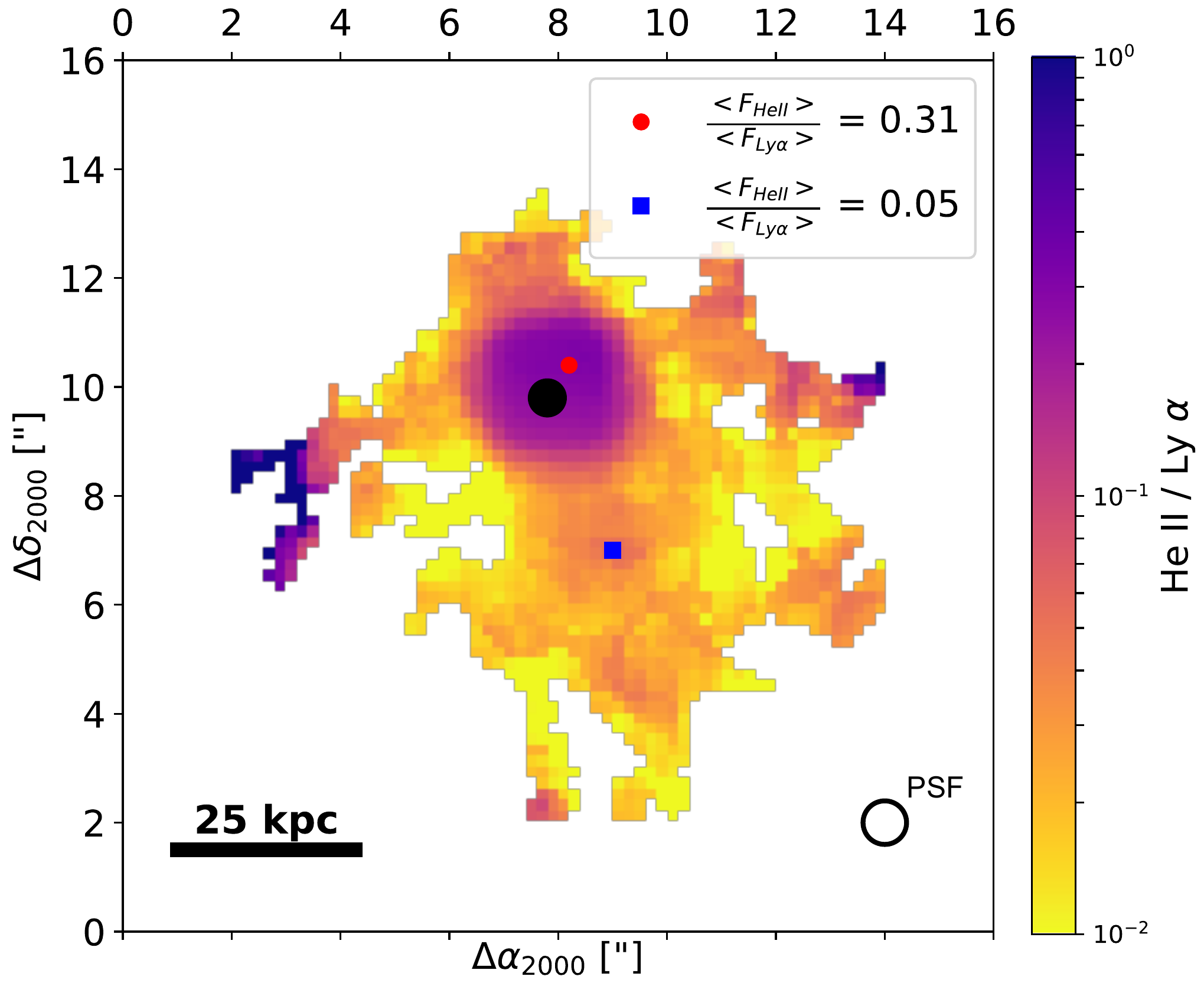}
    \caption{Measured spatial distribution of the \heii\ to Ly$\alpha$ line ratio for the Bulb nebula. The map was generated by using the intersection of the \texttt{CubEx} Ly$\alpha$ and \heii\ 3D segmentation maps where the line emission has S/N$>$2. To maximize the S/N, we used the optimally extracted surface brightness (SB) maps of the Ly$\alpha$ and \heii\ extended emission. The two points indicate the line ratio in the central region as well as in a region further out. The value of $0.31$ is very close to the expected value produced by recombination radiation when both Helium and Hydrogen are highly ionized (or doubly ionised, in the case of helium) as discussed in \citep{Cantalupo2019}. }
    \label{fig:line_ratios}
\end{figure}
%
\autoref{fig:line_ratios} presents the derived two-dimensional \heii\ to Ly$\alpha$ ratio for the Bulb nebula. This ratio can also be used to constrain the origin and physical properties of the emitting gas \citep[see for example][]{Dey2005,Villar2007,Humphrey2008,Battaia2015, Cantalupo2019, Marino2019}. 
We produced this map using the same procedure described in \cite{Cantalupo2019} and \cite{Marino2019}, i.e., we took the intersection of the Ly$\alpha$ and \heii\ segmentation maps given by the \texttt{CubEx} routine. 
Using a squared aperture of $3\times3$ pixels ($0.6\arcsec\times0.6 \arcsec$), we highlight in the figure two positions representative of high ratios (0.3, close to the center) and low ratios (0.05, in the periphery).
The large value of $0.31$, close to the recombination case for fully ionized hydrogen and helium is suggestive of a high ionisation parameter. Further out, where the emission is more asymmetric, line ratios
are closer to the values inferred for other nebulae around type I AGNs, e.g., the Slug \citep{Cantalupo2019}.

\subsection{Kinematics}
As a possible way to compare the properties of the different samples, we have also looked into the resolved kinematics maps of the Ly$\alpha$ nebulae and, for the Bulb, of the \heii\ emission using the distribution of the moments as usually done in the literature (see e.g., \citealt{Marino2019} and references therein).  By comparing the first and second moment maps with the sample of \cite{Borisova2016} we find no significant differences between the type I and type II AGNs. In particular, in both classes of AGN we find velocity shears ranging from around $-300$ km s$^{-1}$ to  $300$ km s$^{-1}$ without clear kinematic patterns or strong velocity gradients. Similarly, the FWHM estimated from the second moments are within the range $400 - 800$ km s$^{-1}$ in both sample.  We will provide a detail study of the kinematics, particularly from the \heii\ emission in a future work. 

\section{Discussion}
\label{sec_disc}
\begin{figure*}
    \centering
    \includegraphics[width = \textwidth]{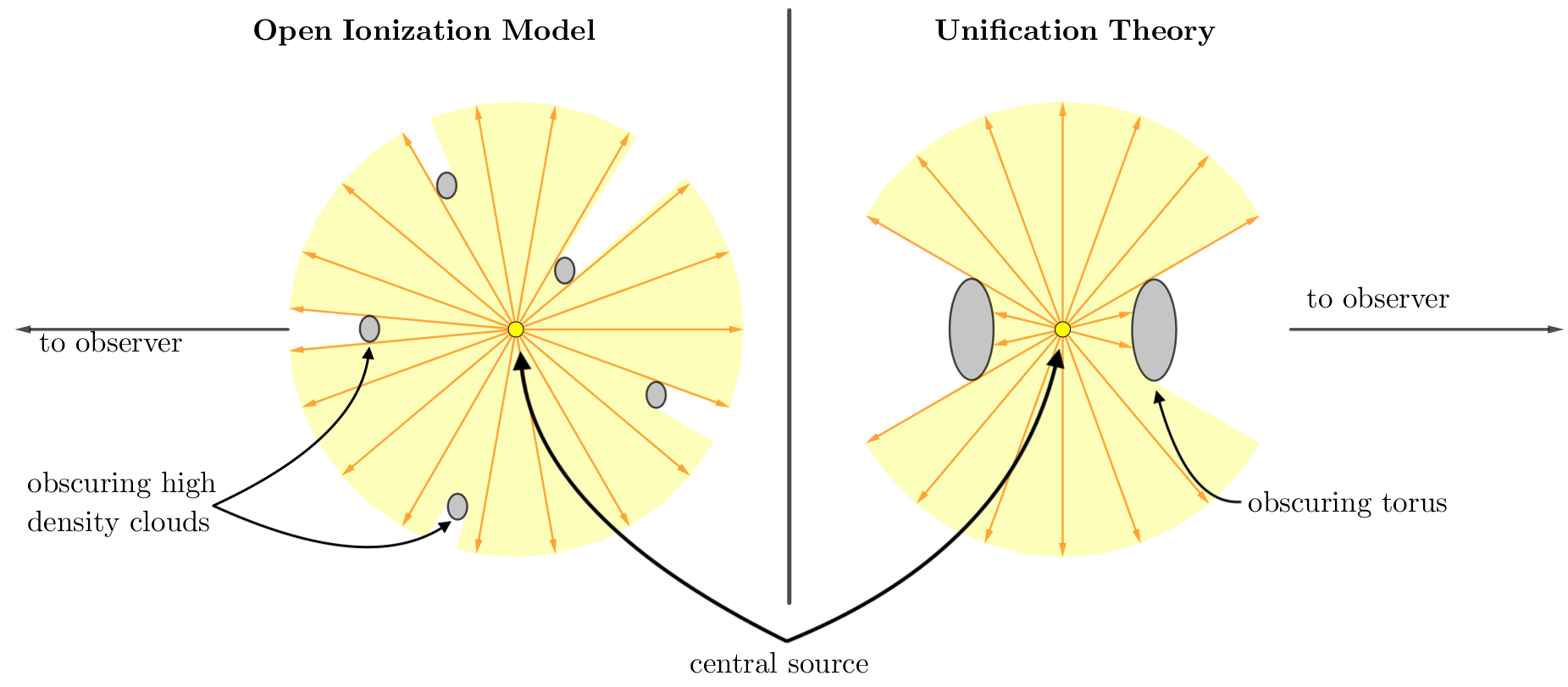}
    \caption{A sketch of two models explaining why AGNs can be highly obscured along our line-of-sight. Left-hand panel, Open Ionisation Model: ionizing radiation from the central source escapes in all direction. High density, dusty clouds obscure the nucleus for an observer from certain viewing angles, making it a type II AGN only along particular (random) directions. However, as there is no internal structure, there are no collimated and clearly defined ionisation cones. Right-hand panel, Unification Theory Model: the central source is obscured along the observer line-of-sight due to a dusty, optically thick torus. The torus is circumnuclear and, as a result, clear ionisation cones are produced. These ionisation structures would then be particularly visible on the plane of the sky for a AGN obscured along our line-of-sight. }
    \label{fig:two_models}
\end{figure*}

The classical AGN unified model attempt to explain certain types of observed properties of AGNs, such as obscuration and structure only by different viewing angles \citep{Antonucci1993, Urry1995}.
According to this model, the gas around the AGN will be ionized along and inside of an ionisation cone, as large parts of the  surrounding gas are shielded from hard ionizing radiation from the central source by an optically thick torus (see \autoref{fig:two_models}, right image). 
While evidence is uncountable for the validity of the unification model in the low redshift Universe \citep[e.g.][]{Schmitt1996,Jaffe2004,Afanasiev2007, Muller2011,Fischer2013, Durre2018} it is not clear wether this also holds in the high redshift Universe. 
For instance, an alternative scenario could be that the obscuration is just due to high columns of gas and dust along the line of sight produced by randomly distributed obscuring material rather than due to a torus (see \autoref{fig:two_models}). 
However, of these two pictures, only the unified model will generate clear ionisation cones. These ionisation cones would then be oriented differently in the type I and type II AGN cases with respect to our line of sight, producing noticeable differences in the asymmetry of the ``illumination" and
therefore of the nebulae. This is instead not expected in the alternative scenario 
\citep{Marino2019}. However, it is also possible that individual clouds in a clumpy medium are responsible for the observed absorption (see \citealt{Marino2019} for an example), rather than a contiguous torus (see \autoref{fig:two_models} for a sketch of the two possibilities). 

Despite the limited statistics, our results, based on two independent methods,  indicate that the nebulae of type II AGN are marginally more asymmetric than their type I counterparts.  In particular, using a moment-based asymmetry indicator, the median of type II AGNs is below the 25 percentile of the type I distribution. A $p$-value of $0.097$ indicates that we have only a 10\% chance, 
that type I and type II nebula arise from the same population in terms of asymmetries. A second method based on  Fourier decomposition (see  \autoref{fig:Fourier_decomp_result}) shows that the type II AGNs  have a more asymmetric spatial morphology, especially at larger distances ($r\gtrsim 40$ pkpc; here the p-value is as low as 5\% is some regions). 
Some nebulae, such as the UDF 09, are even more asymmetric at lower radii ($r\gtrsim 10$ pkpc). Apart from the different degree of asymmetry, the type II nebulae are overall similar to their type I counterparts in terms of overall luminosities and SB local values in function of the distance from the sources, suggesting that the environment and 
CGM properties of these systems (in terms, e.g., of gas densities) is not too dissimilar. We do not think therefore that the statistical differences in asymmetry could be caused by different gas distributions on large scales, although this is certainly possible in individual cases as we see in our larger quasar nebulae sample. Given instead that the obvious differences in these systems
is the AGN class, we think that it is more likely that these are indeed due to the different geometry of the illumination. In this case, our results, at least on large scales, support the idea that an optically thick, dusty torus causes an anisotropic conical shaped ionizing radiation pattern, as suggested by the AGN unification model. 


If our interpretation is correct, the lack of asymmetries on smaller scales (within about 20 kpc from the AGN) necessarily implies additional or different ionisation or emission mechanisms, independent of the presence of the AGN ionisation cones, in the inner regions of the nebulae. 
Note that these regions are much larger than those that could be affected by PSF smoothing, as demonstrated in Fig. D1 in Appendix D.
In the case of Ly$\alpha$ radiation, 
additional emission mechanisms could include "continuum-pumping"/"scattering" or "collisional excitation"/"Ly$\alpha$ cooling"  (see \citet{Cantalupo2017} for a review).
 In both cases, non-resonant lines typically produced by recombinations are needed to exclude these possibilities, e.g. \heii\ $\lambda$1640.
Likely due to its faintness, we have been only able to detect \heii\ $\lambda$1640 in the deepest of our observations around type II AGNs, i.e. in the Bulb nebula. At least for this nebula, we observe a \heii\ to Ly$\alpha$ ratio in the central region that matches very well the expectations from recombination radiation
suggesting that the Ly$\alpha$ emission in the center is not due to additional emission mechanism. This implies that additional ionisation sources, ``illuminating'' the gas also outside of the type II AGN ionizing cones are needed. In addition, these sources should have a spectrum
that is hard enough to doubly ionise helium. Interestingly, recent results obtained with MUSE do show the presence of  \heii\ emission in galaxies at redshifts $z \sim 2-4$ that are not clearly hosting an AGN \citep{Nanayakkara2019}. 
The helium ionizing photons for this case could be
produced by a young and vigorous star formation activity or simply light echoes from a previous AGN episode. It is interesting to notice that the AGN and such high star formation activity could be coexisting in at least some sources in our sample. Alternatively, the emission could also be due
to ionisation produced by recombination radiation in the AGN cones in the inner parts of the nebula or in the host galaxy ISM. Without a detail model of the gas distribution on these small scales it is however difficult with current data to provide a constraint on this possible emission mechanism and future studies of
H$\alpha$ emission at high spatial resolution, e.g. with JWST, would be important to address this issue.



        
\section{Summary}
\label{sec_conc}

We have used the spatial morphology of extended Ly$\alpha$ nebulae detected around high-redshift AGNs
to test the validity of the AGN unification model at $z\sim3$. In particular, 
we have compared the properties of four Ly$\alpha$ nebulae detected around
type II AGNs in different MUSE surveys with 
a sample of 19 previously detected giant Ly$\alpha$ nebulae around type-I AGNs
\citep{Borisova2016}. 

In order to study and quantify the morphologies of the Ly$\alpha$ nebulae we have used two
independent methods based on the second-order moments of the spatial
distribution of the emission and on Fourier decomposition. 
With both methods, we find hints that type II AGN nebulae are marginally more spatially asymmetric
than their type I counterparts, especially at distances larger than 20-30 physical
kpc from the AGN. This is also consistent with a visual inspection of the nebulae that
show triangular structures (``ionisation-cone'' like) for type II AGNs, in some cases.
Overall, the nebulae show similar luminosities and SB profiles across the different
samples, suggesting that the different morphology is indeed related to 
different ``illumination'' geometries, as implied by the AGN unification model. 

The central and brightest regions of the type II AGN nebulae show instead a more circular
morphology. Given the constraints on the large scales and the presence, at least
in one case, of \heii\ emission with \heii\ to Ly$\alpha$ ratios compatible with recombination
radiation, we conclude that additional (and fainter) ionisation sources unrelated to the AGN
should be present. In addition, these sources should have a spectrum
that is hard enough to doubly ionise helium. 

Despite the currently limited sample of type II AGNs, we have shown that by using carefully designed methods 
it is possible to constrain the differences in the spatial morphology between nebulae around different AGN samples
at statistically significant levels. Therefore, we are confident that future larger samples of type II AGN nebulae,
together with the methods (further) developed here, will provide much stronger constraints on the
detailed illumination properties of high-redshift AGNs.



\section*{Acknowledgements}

SC and GP gratefully acknowledge support from Swiss National Science Foundation grant PP00P2\_163824.
MK acknowledges support by DLR500R1904. 
\bibliographystyle{mnras}
\bibliography{references}


\appendix
\section{Observational Data}
\begin{figure*}
    \centering
    \includegraphics[width = 0.9\textwidth]{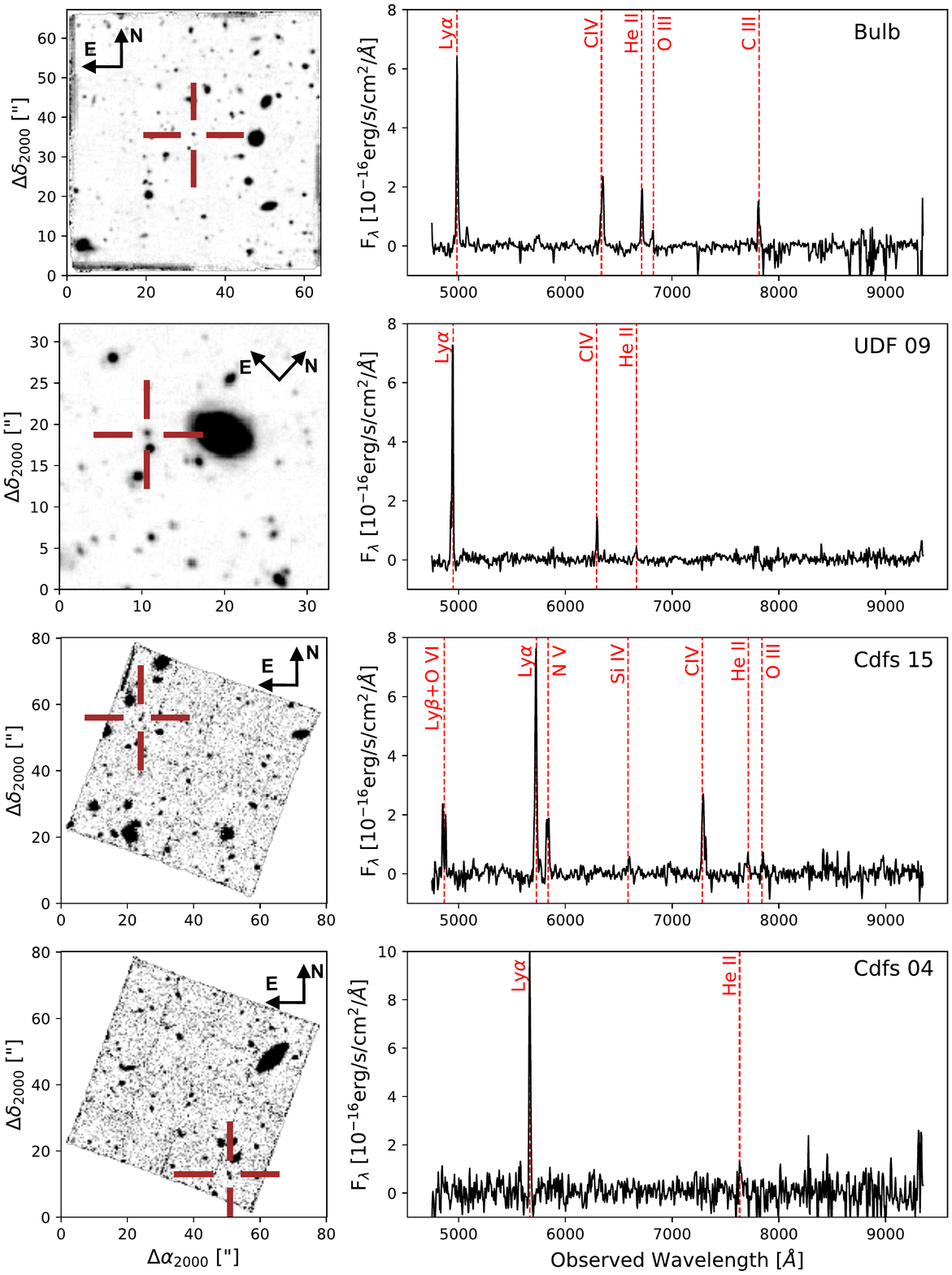}
    \caption{White-light images of the fields around the four type II AGNs in our sample, obtained by collapsing the MUSE cube along the spectral direction. The cross indicates the location of the AGN. The spectra are obtained by integrating within an aperture defined by the spatial (2D) projection of the 3D segmentation map obtained by CubEx (see text for details). The most relevant emission lines are indicated. 
    }
    \label{fig:White_Light}
\end{figure*}

In \autoref{fig:White_Light} the white light images of the fields of the type II AGNs studied in this work are shown together their integrated spectrum.  The spectra are obtained by integrating using an aperture defined by the spatial, 2D projection of the 3D segmentation map obtained by CubEx as described in the main text. 

\section{Pseudo Narrow--Band Images}
\begin{figure*}
    \centering
    \includegraphics[width = 0.9\textwidth]{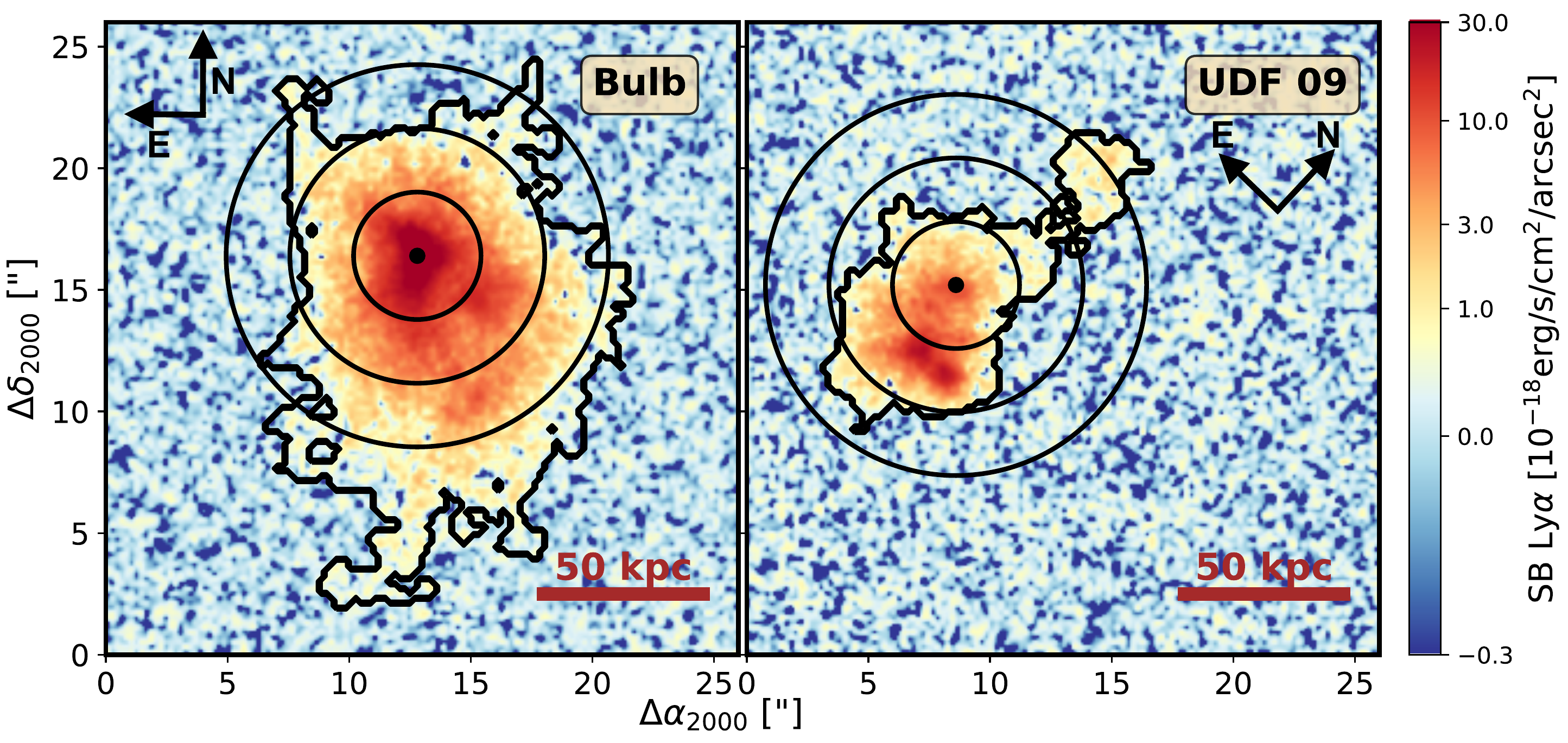}
    \caption{Pseudo-narrow-band Ly$\alpha$ images of the 4 detected nebulae in this study. The wavelength integration window has been set by the maximum width of the 3D segmentation map obtained by CubEx. Annuli are centered on the AGN position and have radii of 20, 40 and 60 physical kpc.}
    \label{fig:NB_images}
\end{figure*}
\begin{figure*}
    \centering
    \includegraphics[width = 0.9\textwidth]{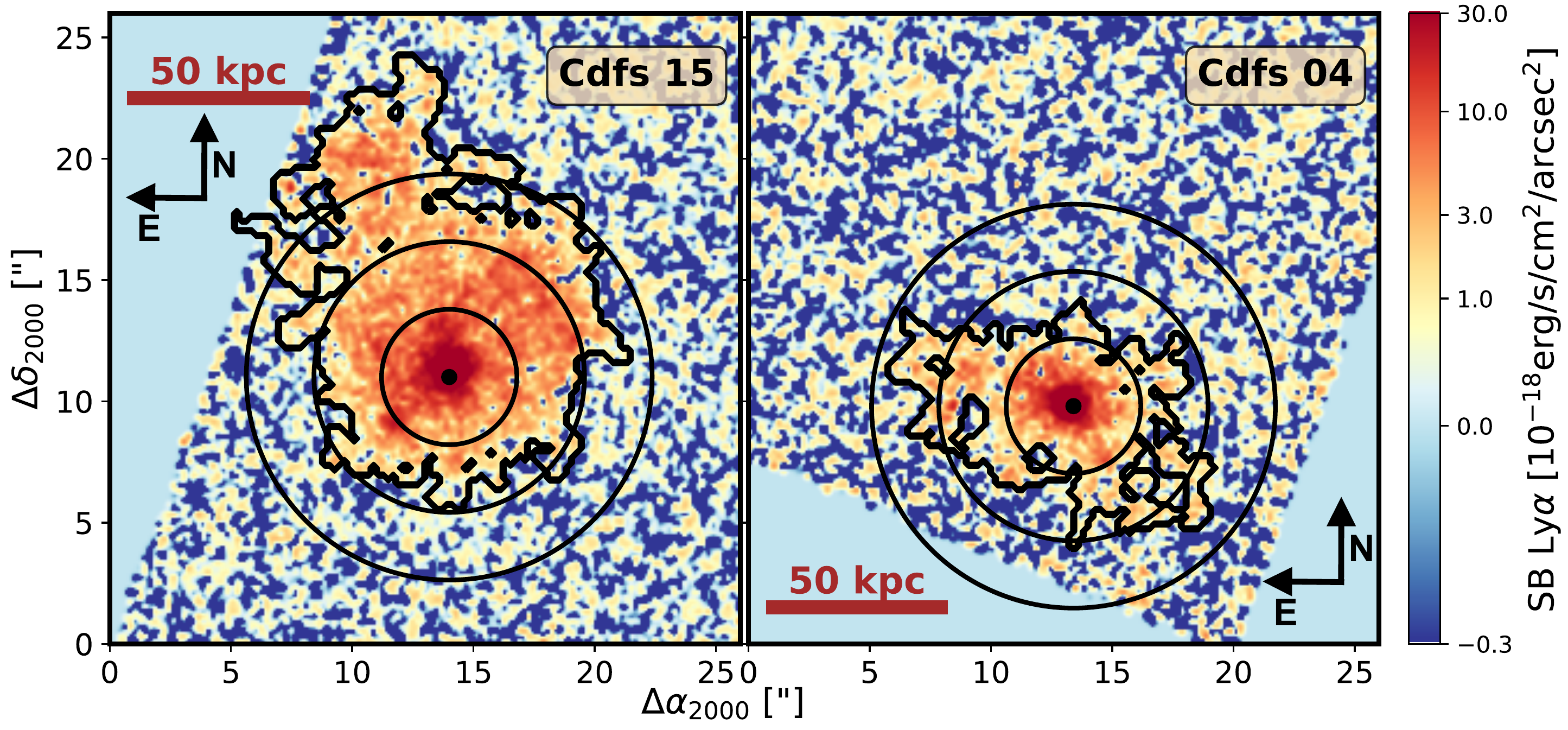}
    \caption{Pseudo-narrow-band Ly$\alpha$ images of the 4 detected nebulae in this study. The wavelength integration window has been set by the maximum with of the 3D segmentation map obtained by CubEx.}
    \label{fig:NB_images_2}
\end{figure*}

\autoref{fig:NB_images} and \autoref{fig:NB_images_2} show the pseudo narrow-band images of the Ly$\alpha$ nebulae detected around the four type II AGNs in this study. The wavelength integration window has been chosen using the maximum width of the 3D segmentation map obtained by CubEx as described in the main text. The concentric rings show distances of 20, 40 and 60 physical kpc from the AGN. 




\section{Surface Brightness Profile Fitting}
\begin{figure*}
    \centering
    \includegraphics[width = 0.75\textwidth]{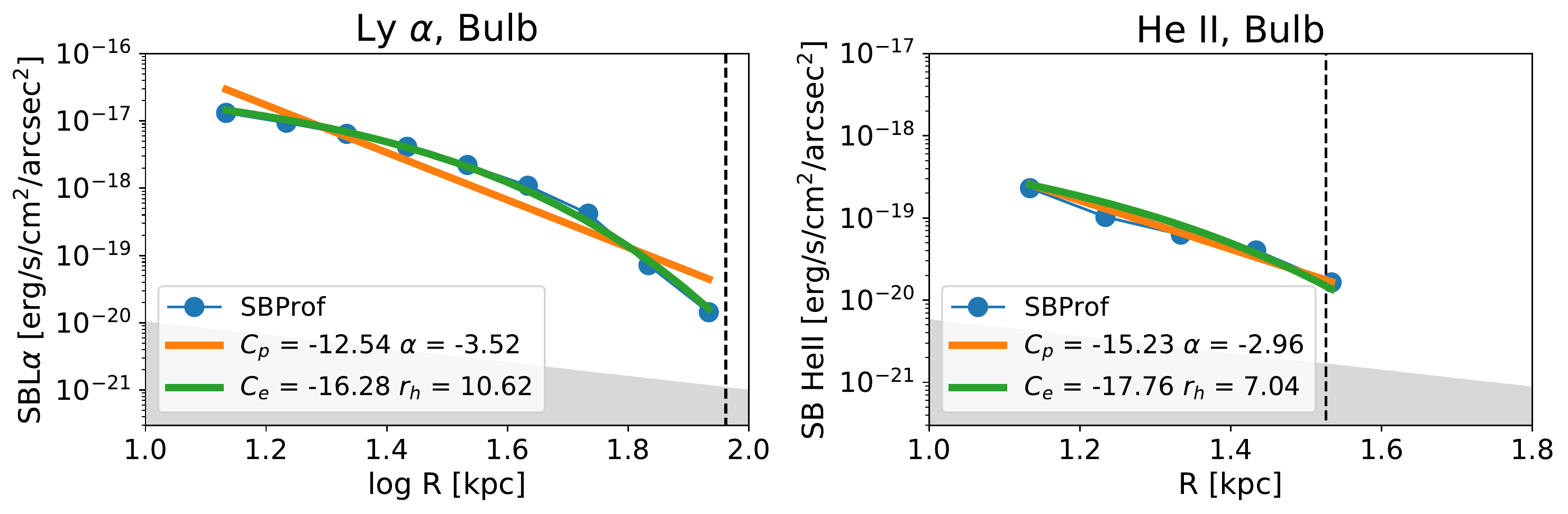}
    \includegraphics[width = 0.37125\textwidth]{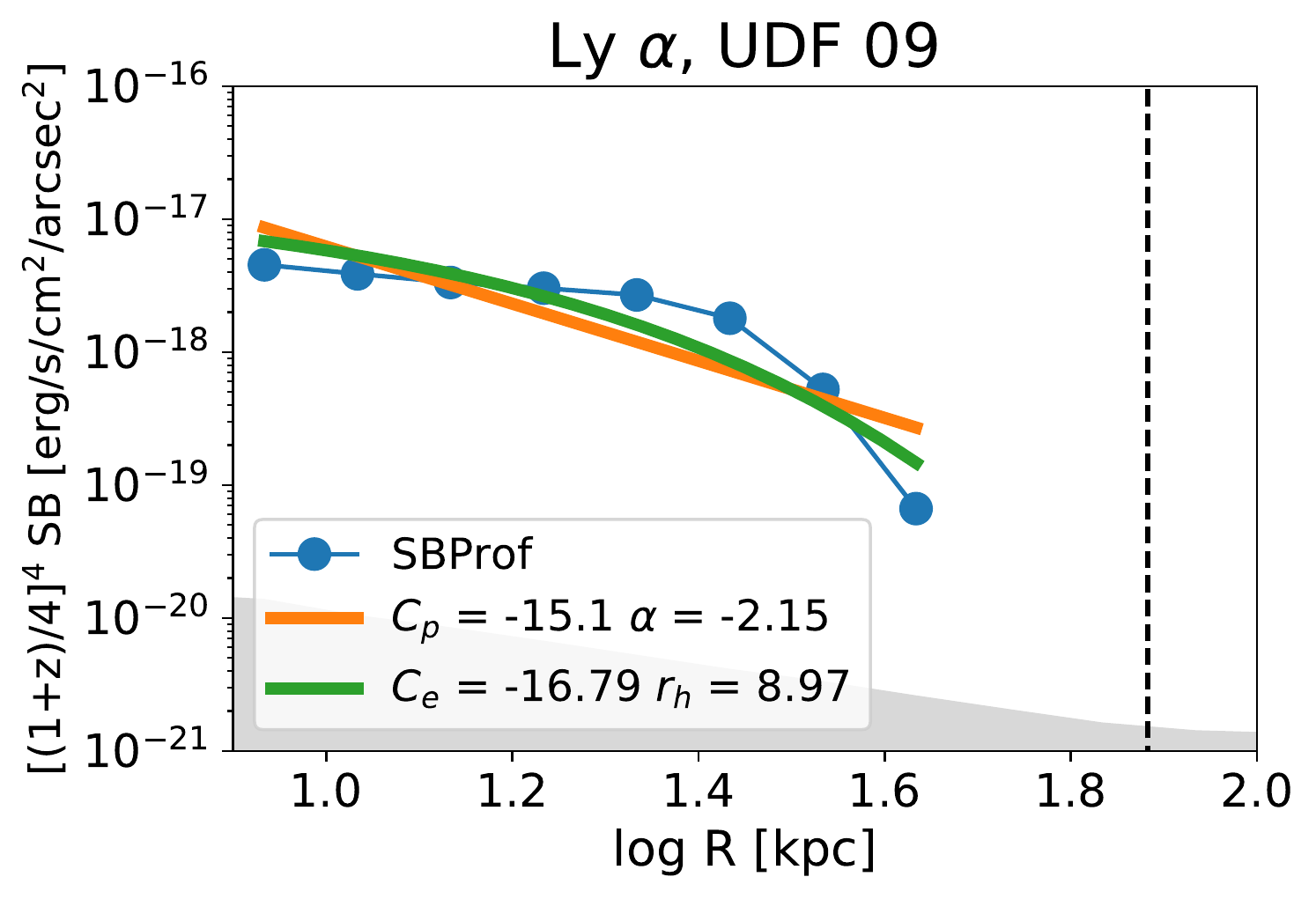}
    \includegraphics[width = 0.37125\textwidth]{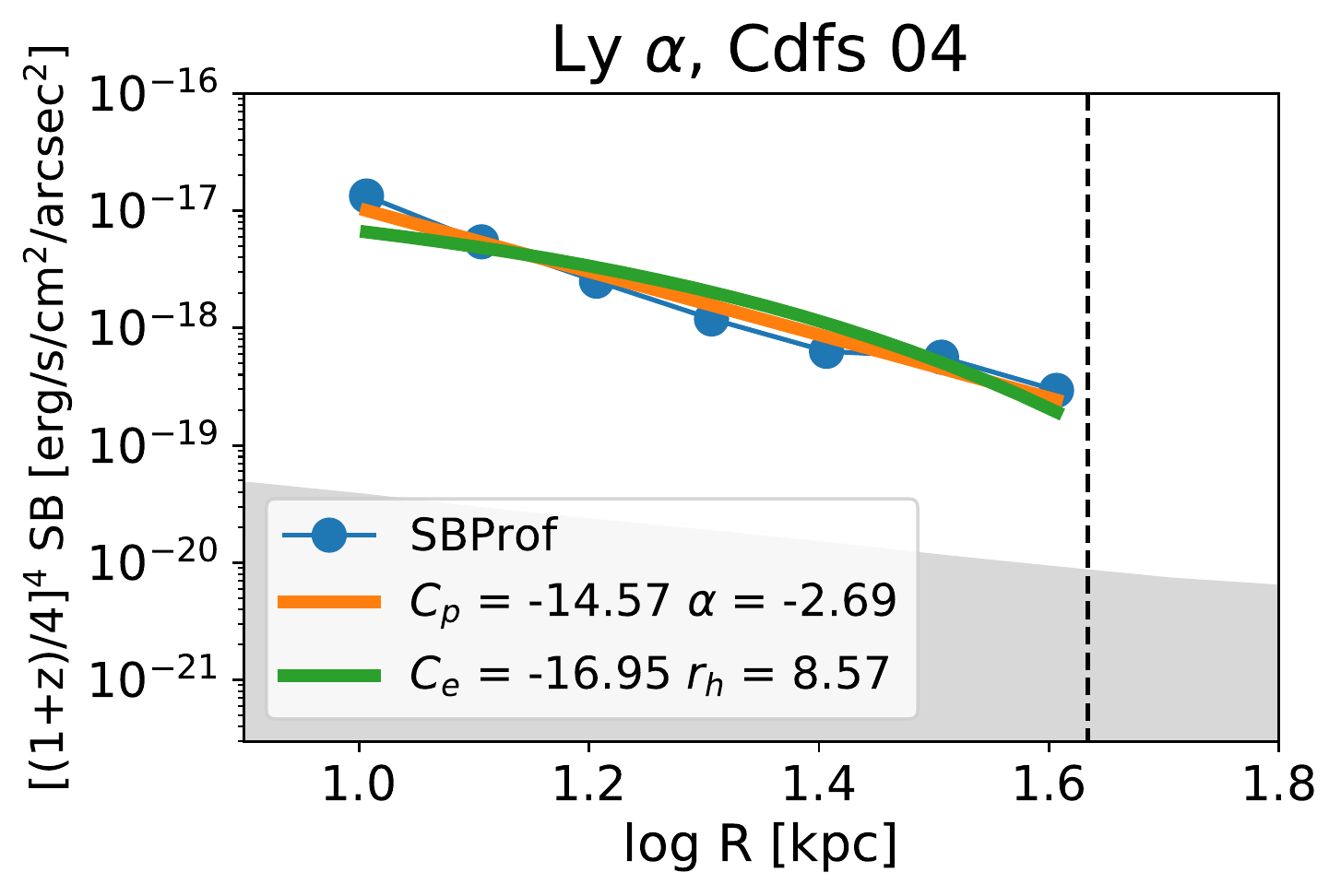}
    \includegraphics[width = 0.37125\textwidth]{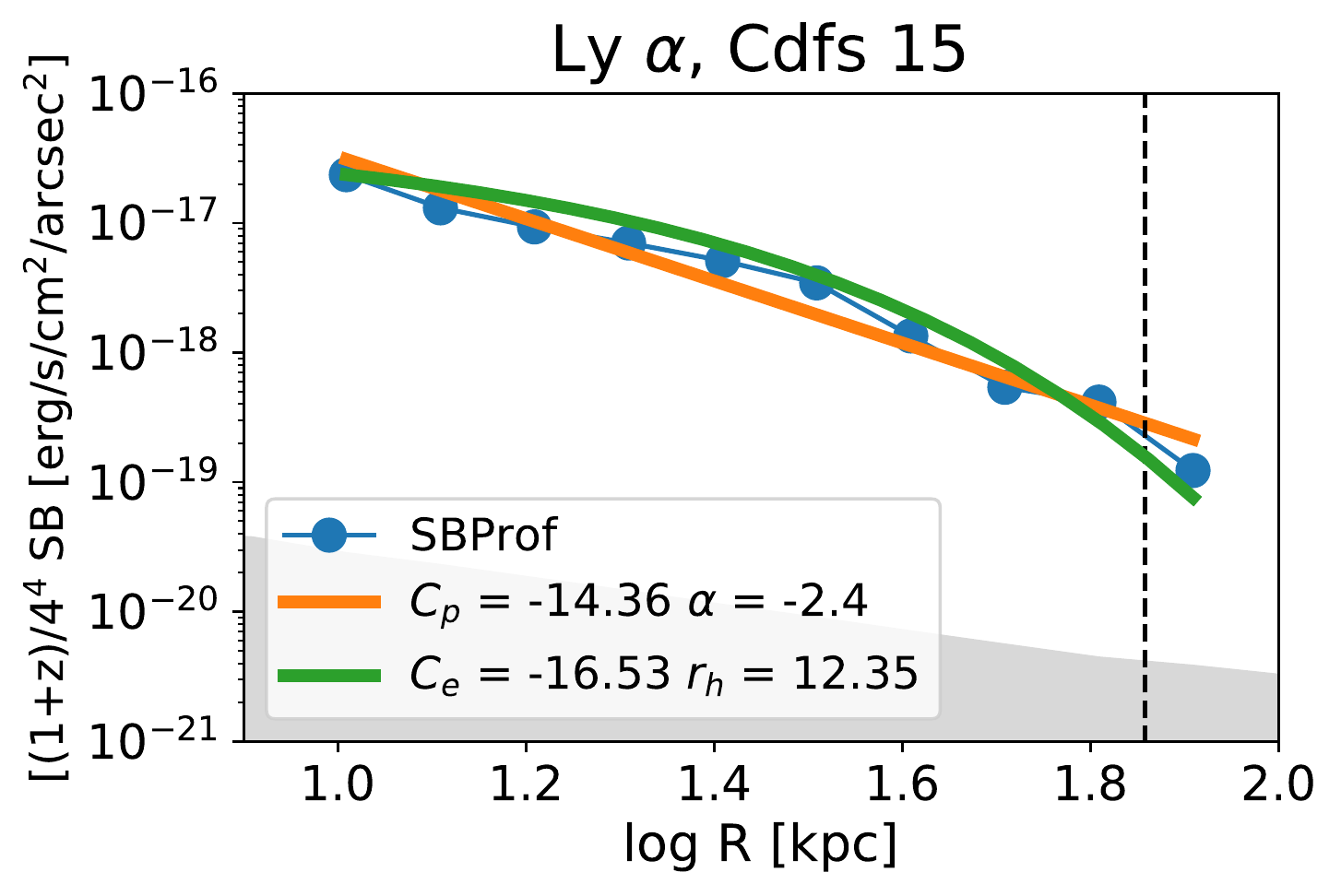}
    \caption{Individual circularly-averaged Ly$\alpha$ (and \heii\ $\lambda$1640) SB profiles for the four type II AGNs in our sample (blue dots with lines). Fitted power-law (orange) and exponential (green) profiles are overlaid. The vertical black striped line indicates the maximal extension of the nebula. The shaded area indicates the noise level for each bin.}
    \label{fig:SB_profiles}
\end{figure*}

We fitted two models to the surface brightness profiles: a power law and and exponential analytical profile.
The power law is described by:
$$
    SB(r) = C_pr^\alpha
$$
and the exponential profile is described by:
$$
SB(r) = C_ee^{-r/r_h}.
$$
\autoref{tab:Result_Fitting} summarizes the fitting procedure results and the \autoref{fig:SB_profiles} show the individual surface brightness profiles.

\begin{table*}
            \centering
            \caption[Result Fitting]{Results from SB profile fitting routine.}
        \label{tab:Result_Fitting}
        \begin{tabular}{c c c c c c }
             \hline
            Name &Line& $\alpha^\text{a}$ & $\log(C_p)^{\text{b}}$ & $r_h^\text{c}$  & $\log(C_e)^{\text{b}}$ \\
              & & &  &[pkpc]&\\ \hline \hline
             \multirow{2}{*}{Bulb} & Ly$\alpha$& $-3.52$ & $-12.54$&$10.62$&$-16.28$\\
             & \heii\ & $-2.96$ & $-15.23$&$7.04$&$-17.76$\\ \hline
             UDF 09 & Ly$\alpha$& $-2.15$ & $-15.1$&$8.97$&$-16.79$\\ \hline
             Cdfs 04 & Ly$\alpha$& $-2.69$ & $-14.57$&$8.57$&$-16.95$\\ \hline
             Cdfs 15 & Ly$\alpha$& $-2.40$ & $-14.36$&$12.35$&$-16.53$\\
            
        \end{tabular}
        \vspace{0.3 cm}
    
        \raggedright
        
        { \footnotesize Notes.\\$^{\text{a}}$ Power-law fit slope to the surface brightness profile.  \\
        $^{\text{b}}$  Normalization parameter for the power-law\\
        $^{\text{c}}$ Scale length from the exponential-law fit \\
        $^{\text{d}}$ Normalization parameter for the exponential-law.\\
        
        }
    
\end{table*}

\section{PSF Effects on the Extended Line Emission}
\begin{figure}
	 \centering
    	\includegraphics[width = 0.9\columnwidth]{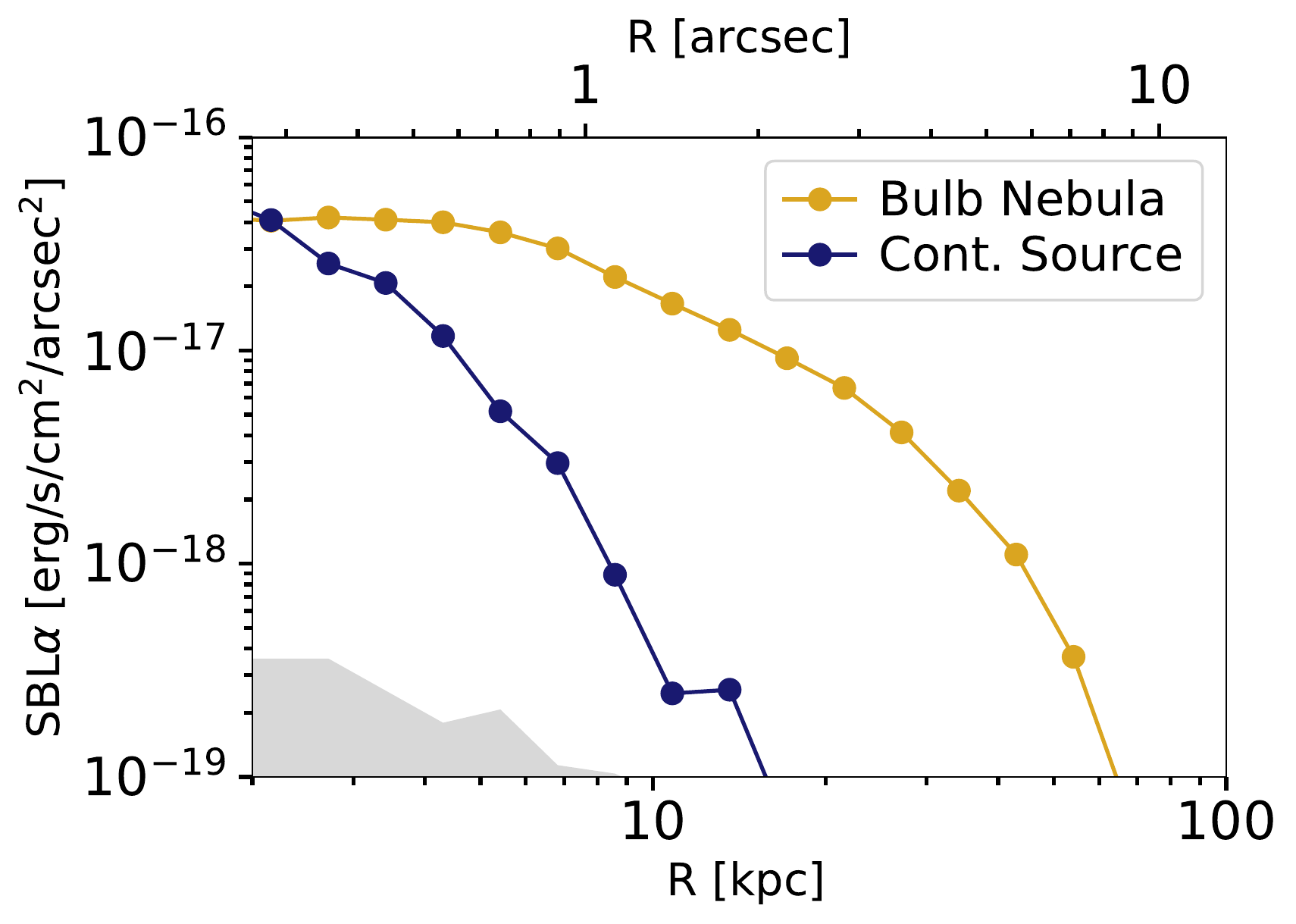}
    	\caption{The Ly$\alpha$ surface brightness (SB) profile using the pseudo narrow-band (pseudo NB) MUSE image. The golden line is the circularly averaged profile of the bulb extended Ly$\alpha$ emission. The purple line indicates the profile of a continuum source. This SB profile has been renormalized such that the most central surface brightness corresponds to the bulb nebula's most central surface brightness. The shaded area indicates the noise level for each bin.}
    \label{fig:SBProf_comp}
\end{figure}

To study PSF effects we investigate the SB profile of the extended Ly$\alpha$ emission of the bulb nebula with the SB profile of a continuum source. \autoref{fig:SBProf_comp} compares the two SB profiles. The continuum SB is computed from the same MUSE cube. We took the AGN continuum emission and selected a wavelength window starting at 6000 $\AA$ with the same width as used for the pseudo NB Ly$\alpha$ image. We see that for $R>5$ kpc the PSF SB is less than 10\% of the Ly$\alpha$ SB.



\label{lastpage}
\end{document}